\documentclass{article} 
\usepackage[margin=1in]{geometry}
\usepackage{times}
\usepackage{natbib}
\usepackage{parskip}

\usepackage{amsmath,amsfonts,bm}

\def\eqref#1{equation~\ref{#1}}

\def\1{\bm{1}}

\DeclareMathAlphabet{\mathsfit}{\encodingdefault}{\sfdefault}{m}{sl}
\SetMathAlphabet{\mathsfit}{bold}{\encodingdefault}{\sfdefault}{bx}{n}

\DeclareMathOperator*{\argmax}{arg\,max}

\usepackage{xcolor}
\usepackage[normalem]{ulem} 
\usepackage{todonotes}

\newif\ifhideoldsuggestions

\definecolor{suggest0}{RGB}{100,53,69}
\definecolor{suggest1}{RGB}{140,53,69}
\definecolor{suggest2}{RGB}{180,53,69}
\definecolor{suggest3}{RGB}{220,53,69}
\definecolor{suggest4}{RGB}{40,167,69}

\makeatletter
\newif\ifsuggest@floats
\newcommand{\suggest@floats}[1]{%
  \def\suggest@floatcolor{#1}%
  \ifsuggest@floats\else
    \suggest@floatstrue
    \expandafter\def\expandafter\@floatboxreset\expandafter{\@floatboxreset\color{\suggest@floatcolor}}%
    \expandafter\def\expandafter\@endfloatbox\expandafter{\@endfloatbox\suggest@tint}%
  \fi}
\newcommand{\suggest@tintsep}{3pt}
\newcommand{\suggest@tint}{%
  \global\setbox\@currbox\vbox{\hbox{%
    \rlap{\kern-\suggest@tintsep\smash{\color{\suggest@floatcolor!10}\vrule
      \@width\dimexpr\wd\@currbox+\suggest@tintsep+\suggest@tintsep\relax
      \@height\dimexpr\ht\@currbox+\suggest@tintsep\relax
      \@depth\dimexpr\dp\@currbox+\suggest@tintsep\relax}}%
    \box\@currbox}}}

\ExplSyntaxOn
\seq_new:N \l__suggest_versions_seq
\seq_new:N \l__suggest_pars_seq
\tl_new:N \l__suggest_chunk_tl
\int_new:N \l__suggest_count_int
\int_new:N \l__suggest_depth_int
\int_new:N \l__suggest_tmp_int

\NewDocumentCommand \suggest { }
  {
    \seq_clear:N \l__suggest_versions_seq
    \__suggest_collect:
  }

\cs_new_protected:Npn \__suggest_collect:
  {
    \peek_catcode:NTF \c_group_begin_token
      { \__suggest_grab:n }
      { \__suggest_render: }
  }

\cs_new_protected:Npn \__suggest_grab:n #1
  {
    \seq_put_right:Nn \l__suggest_versions_seq {#1}
    \__suggest_collect:
  }

\cs_new_protected:Npn \__suggest_render:
  {
    \int_set:Nn \l__suggest_count_int { \seq_count:N \l__suggest_versions_seq }
    \seq_map_indexed_inline:Nn \l__suggest_versions_seq
      {
        \int_compare:nNnTF {##1} < { \l__suggest_count_int }
          { \legacy_if:nF { hideoldsuggestions } { \__suggest_old:nn {##1} {##2} } }
          {
            \legacy_if:nTF { hideoldsuggestions }
              {##2}
              { \__suggest_latest:nn {##1} {##2} }
          }
      }
  }

\NewDocumentCommand \__suggest_boxed_cite:N { m o o m }
  {
    \mbox
      {
        \IfNoValueTF {#2} { #1 {#4} }
          { \IfNoValueTF {#3} { #1 [#2] {#4} } { #1 [#2] [#3] {#4} } }
      }
  }
\cs_new_protected:Npn \__suggest_protect_cites:
  {
    \clist_map_inline:nn { citep, citet, cite, citealp, citealt, citeauthor, citeyear, citeyearpar }
      {
        \cs_if_exist:cT { ##1 }
          {
            \cs_set_eq:cc { __suggest_orig_ ##1 : } { ##1 }
            \cs_set:cpx { ##1 } { \exp_not:N \__suggest_boxed_cite:N \exp_not:c { __suggest_orig_ ##1 : } }
          }
      }
  }

\cs_new_protected:Npn \__suggest_old:nn #1#2
  {
    \group_begin:
    \__suggest_protect_cites:
    \__suggest_if_block:nTF {#2}
      {
        \__suggest_color_block:nn {#1}
          {
            \int_compare:nNnT { \l__suggest_count_int } > { 2 }
              { \sout { \__suggest_prefix:n {#1} } }
            \__suggest_strike_pars:n {#2}
          }
      }
      {
        \__suggest_color:nn {#1} { \sout { \__suggest_prefix:n {#1} #2 } }
        \hspace { 0.2em }
      }
    \group_end:
  }

\cs_new_protected:Npn \__suggest_strike_pars:n #1
  {
    \seq_set_split:Nnn \l__suggest_pars_seq { \par } {#1}
    \tl_clear:N \l__suggest_chunk_tl
    \int_zero:N \l__suggest_depth_int
    \seq_map_inline:Nn \l__suggest_pars_seq
      {
        \tl_if_empty:NF \l__suggest_chunk_tl { \tl_put_right:Nn \l__suggest_chunk_tl { \par } }
        \tl_put_right:Nn \l__suggest_chunk_tl {##1}
        \regex_count:NnN \c__suggest_begin_regex {##1} \l__suggest_tmp_int
        \int_add:Nn \l__suggest_depth_int { \l__suggest_tmp_int }
        \regex_count:NnN \c__suggest_end_regex {##1} \l__suggest_tmp_int
        \int_sub:Nn \l__suggest_depth_int { \l__suggest_tmp_int }
        \int_compare:nNnF { \l__suggest_depth_int } > { 0 } { \__suggest_strike_chunk: }
      }
    \__suggest_strike_chunk:
  }

\cs_new_protected:Npn \__suggest_strike_chunk:
  {
    \tl_if_blank:VF \l__suggest_chunk_tl
      { \exp_args:NV \__suggest_strike_par:n \l__suggest_chunk_tl }
    \tl_clear:N \l__suggest_chunk_tl
    \int_zero:N \l__suggest_depth_int
  }

\cs_new_protected:Npn \__suggest_strike_par:n #1
  { \__suggest_if_block:nTF {#1} {#1} { \sout {#1} } \par }

\cs_new_protected:Npn \__suggest_latest:nn #1#2
  {
    \__suggest_if_block:nTF {#2}
      { \__suggest_color_block:nn {#1} { \__suggest_prefix:n {#1} #2 } }
      { \__suggest_color:nn {#1} { \__suggest_prefix:n {#1} #2 } }
  }

\regex_const:Nn \c__suggest_block_regex
  { \c{par|begin|item|(sub)*section|(sub)?paragraph|\[} }
\regex_const:Nn \c__suggest_begin_regex { \c{begin} }
\regex_const:Nn \c__suggest_end_regex { \c{end} }

\cs_new_protected:Npn \__suggest_if_block:nTF
  { \regex_match:NnTF \c__suggest_block_regex }

\cs_new:Npn \__suggest_color_name:n #1
  { suggest \int_max:nn { 0 } { 4 - \l__suggest_count_int + #1 } }

\cs_new_protected:Npn \__suggest_color:nn #1#2
  { \exp_args:Ne \textcolor { \__suggest_color_name:n {#1} } {#2} }

\cs_new_protected:Npn \__suggest_color_block:nn #1#2
  {
    \group_begin:
      \exp_args:Ne \color { \__suggest_color_name:n {#1} }
      \exp_args:Ne \suggest@floats { \__suggest_color_name:n {#1} }
      #2
    \group_end:
  }

\cs_new_protected:Npn \__suggest_prefix:n #1
  { \int_compare:nNnT { \l__suggest_count_int } > { 2 } { v #1 :~ } }
\ExplSyntaxOff
\makeatother

\providecommand{\codingpromptsize}{\small}
\providecommand{\codingpromptskip}{0.6\baselineskip}
\newenvironment{codingprompt}{\par\begingroup\codingpromptsize\ttfamily\raggedright\frenchspacing
  \setlength{\parindent}{0pt}\setlength{\parskip}{\codingpromptskip}\setlength{\leftskip}{1em}}{\par\endgroup}

\usepackage{hyperref}
\usepackage{url}
\usepackage{bbm}
\usepackage{amssymb,amsmath,amsthm}
\usepackage{mathtools}
\usepackage{booktabs}
\usepackage{tikz}
\theoremstyle{plain}
\newtheorem{theorem}{Theorem}[section]
\newtheorem{proposition}[theorem]{Proposition}

\newtheorem{condition}{Condition}

\definecolor{drpLine}{RGB}{38,38,138}
\definecolor{drpCell}{RGB}{232,240,252}
\definecolor{drpMerged}{RGB}{253,232,205}
\definecolor{drpSame}{RGB}{27,122,62}
\definecolor{drpDiff}{RGB}{110,60,160}
\newcommand{\drpCell}[6]{%
  \filldraw[fill=#1, draw=drpLine, line width=0.4pt] (#2,#3) rectangle ++(#4,#5);
  \node[font=\footnotesize, inner sep=0pt] at (#2+0.5*#4,#3+0.5*#5) {#6};}
\newcommand{\drpFrame}{\draw[drpLine, line width=0.9pt] (0,0) rectangle (2,2);}
\newcommand{\drpGridFour}[4]{%
  \drpCell{drpCell}{0}{1}{1}{1}{#1}\drpCell{drpCell}{1}{1}{1}{1}{#2}%
  \drpCell{drpCell}{0}{0}{1}{1}{#3}\drpCell{drpCell}{1}{0}{1}{1}{#4}\drpFrame}
\newcommand{\drpGridTop}[3]{%
  \drpCell{drpMerged}{0}{1}{2}{1}{#1}%
  \drpCell{drpCell}{0}{0}{1}{1}{#2}\drpCell{drpCell}{1}{0}{1}{1}{#3}\drpFrame}
\newcommand{\drpGridRows}[2]{%
  \drpCell{drpMerged}{0}{1}{2}{1}{#1}\drpCell{drpMerged}{0}{0}{2}{1}{#2}\drpFrame}
\newcommand{\drpGridOne}[1]{%
  \drpCell{drpMerged}{0}{0}{2}{2}{#1}\drpFrame}

\title{Decision-Relevant Predictions:\\Attention Management with Joint Scoring Rules}

\author{Rubi Hudson\thanks{University of Toronto and Principled Agents. Correspondence: \texttt{contact@principledagents.org}} \and Baram Sosis\thanks{Principled Agents} \and Artem Petrov\footnotemark[2]}

\begin{document}

\maketitle

\begin{abstract}
\noindent
When a decision-making principal wishes to elicit predictions from an agent using a proper scoring rule, the principal must specify the outcomes to be assigned probability. This can result in the principal either missing important considerations or being given too much information to process. We would instead like the agent making the predictions to provide the decision-relevant set of outcomes. This could be either the simplest set of outcomes leading the principal to the decision they would make under full information, or the simplest set of outcomes where no further information would change the principal's decision. We show that it is impossible to strictly incentivize a single agent to provide either of these sets, but that it can be done with two or more agents evaluated jointly. This is achieved by rewarding agents for refining the outcome set in a way that changes the principal's decision. We then show how to adapt the mechanisms to elicit a prediction alongside the decision-relevant outcome set, for cases where the set of possible outcomes is very large or continuous. Applications are demonstrated, first with a simple experiment where agents reveal a minimal mask of an MNIST image that allows a judge to classify it, and then by using language models to generate a small set of unit tests that allow a weaker judge to accurately evaluate code.
\end{abstract}

\section{Introduction}
When making a decision between possible courses of action, a common step is to predict the likely outcomes of each. For important decisions outside of our normal expertise, we may consult an expert and use their predictions instead of our own, or alongside it. While honest predictions can be incentivized from an expert using a proper scoring rule \citep{Brier1950,Good1952}, there is no best practice for determining which set of outcomes to use.

A decision maker constructing the outcome set can miss important considerations, leading to suboptimal decisions even after eliciting an expert's honest prediction. However, including all possible details can quickly make predictions too large to evaluate, or even elicit. What we would like instead is for the expert, who knows which outcomes are relevant, to partition the set of outcomes so that it includes all necessary information but no extraneous details.

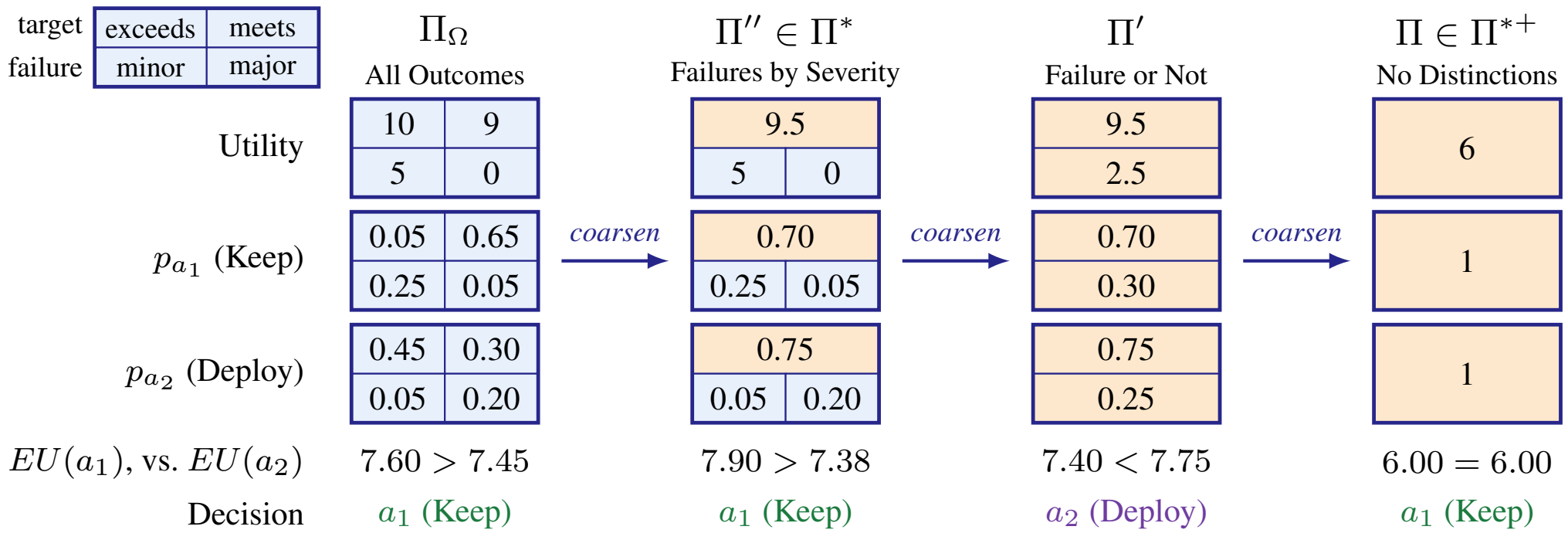
\begin{figure}[t]
\centering
\begin{tikzpicture}[x=0.8cm, y=0.42cm]
  \node[anchor=east, font=\footnotesize] at (-0.35,5.6) {Utility};
  \node[anchor=east, font=\footnotesize] at (-0.35,3.3) {$p_{a_1}$ (Keep)};
  \node[anchor=east, font=\footnotesize] at (-0.35,1.0) {$p_{a_2}$ (Deploy)};
  \node[anchor=east, font=\footnotesize] at (-0.35,-0.8) {$EU(a_1)$,\ vs.\  $EU(a_2)$};
  \node[anchor=east, font=\footnotesize] at (-0.35,-1.85) {Decision};

  \begin{scope}[shift={(-2.75,6.85)}]
    \drpCell{drpCell}{0}{0.8}{1.2}{0.8}{\scriptsize exceeds}
    \drpCell{drpCell}{1.2}{0.8}{1.2}{0.8}{\scriptsize meets}
    \drpCell{drpCell}{0}{0}{1.2}{0.8}{\scriptsize minor}
    \drpCell{drpCell}{1.2}{0}{1.2}{0.8}{\scriptsize major}
    \draw[drpLine, line width=0.9pt] (0,0) rectangle (2.4,1.6);
    \node[anchor=east, font=\scriptsize, inner sep=0pt] at (-0.12,1.2) {target};
    \node[anchor=east, font=\scriptsize, inner sep=0pt] at (-0.12,0.4) {failure};
  \end{scope}

  \begin{scope}[shift={(0,0)}]
    \node[font=\small] at (1,8.0) {$\Pi_\Omega$};
    \node[font=\scriptsize] at (1,7.1) {All Outcomes};
    \begin{scope}[shift={(0,4.6)}]\drpGridFour{10}{9}{5}{0}\end{scope}
    \begin{scope}[shift={(0,2.3)}]\drpGridFour{0.05}{0.65}{0.25}{0.05}\end{scope}
    \begin{scope}[shift={(0,0.0)}]\drpGridFour{0.45}{0.30}{0.05}{0.20}\end{scope}
    \node[font=\footnotesize] at (1,-0.8) {$7.60 > 7.45$};
    \node[font=\footnotesize, text=drpSame] at (1,-1.85) {$a_1$ (Keep)};
  \end{scope}

  \begin{scope}[shift={(3.65,0)}]
    \node[font=\small] at (1,8.0) {$\Pi''\in\Pi^{*}$};
    \node[font=\scriptsize] at (1,7.1) {Failures by Severity};
    \begin{scope}[shift={(0,4.6)}]\drpGridTop{9.5}{5}{0}\end{scope}
    \begin{scope}[shift={(0,2.3)}]\drpGridTop{0.70}{0.25}{0.05}\end{scope}
    \begin{scope}[shift={(0,0.0)}]\drpGridTop{0.75}{0.05}{0.20}\end{scope}
    \node[font=\footnotesize] at (1,-0.8) {$7.90 > 7.38$};
    \node[font=\footnotesize, text=drpSame] at (1,-1.85) {$a_1$ (Keep)};
  \end{scope}

  \begin{scope}[shift={(7.3,0)}]
    \node[font=\small] at (1,8.0) {$\Pi'$};
    \node[font=\scriptsize] at (1,7.1) {Failure or Not};
    \begin{scope}[shift={(0,4.6)}]\drpGridRows{9.5}{2.5}\end{scope}
    \begin{scope}[shift={(0,2.3)}]\drpGridRows{0.70}{0.30}\end{scope}
    \begin{scope}[shift={(0,0.0)}]\drpGridRows{0.75}{0.25}\end{scope}
    \node[font=\footnotesize] at (1,-0.8) {$7.40 < 7.75$};
    \node[font=\footnotesize, text=drpDiff] at (1,-1.85) {$a_2$ (Deploy)};
  \end{scope}

  \begin{scope}[shift={(10.95,0)}]
    \node[font=\small] at (1,8.0) {$\Pi\in\Pi^{*+}$};
    \node[font=\scriptsize] at (1,7.1) {No Distinctions};
    \begin{scope}[shift={(0,4.6)}]\drpGridOne{6}\end{scope}
    \begin{scope}[shift={(0,2.3)}]\drpGridOne{1}\end{scope}
    \begin{scope}[shift={(0,0.0)}]\drpGridOne{1}\end{scope}
    \node[font=\footnotesize] at (1,-0.8) {$6.00 = 6.00$};
    \node[font=\footnotesize, text=drpSame] at (1,-1.85) {$a_1$ (Keep)};
  \end{scope}

  \foreach \xa in {2.25, 5.9, 9.55} {
    \draw[-latex, drpLine, line width=0.8pt] (\xa,3.3) -- ++(1.15,0)
      node[midway, above=1pt, font=\scriptsize\itshape, text=drpLine] {coarsen};
  }
\end{tikzpicture}
\caption{A principal chooses between keeping their old software ($a_1$) or deploying an update ($a_2$), using predictions on up to four outcomes. The principal maximizes their expected utility (EU), splitting probability evenly between any outcomes grouped together, and defaults to the old software in case of a tie. $\Pi^*$ is the set of simplest partitions for which no refinement changes the decision, while $\Pi^{*+}$ is the simplest partition leading to the full information decision.}
\label{fig:toy-example}
\end{figure}

As an example, Figure \ref{fig:toy-example} shows a principal deciding whether to keep an old software system ($a_1$) or deploy a new one ($a_2$). The principal assigns utilities $(10, 9, 5, 0)$ to outcomes (``exceeds target'', ``meets target'', ``minor failure'', ``major failure'') respectively, and makes their choice to maximize expected utility. The expert knows that outcome probabilities given $a_1$ are $p_{a_1} = (0.05, 0.65, 0.25, 0.05)$ and given $a_2$ are $p_{a_2} = (0.45, 0.30, 0.05, 0.20)$.

Coarsening ``exceeds target'' and ``meets target'' into one outcome would not change choice of action, so all decision-relevant information is provided. However, coarsening ``minor failure'' with ``major failure'' as well would lead to a different action, meaning decision-relevant information was lost. For motivation, we can think of outcomes like ``major failure'' as already coarsening over many different technical issues that the principal does not care to distinguish.

The example shows how there are two types of partitions that would be reasonable to elicit. The first is the simplest such that no refinement would lead to a different choice of action, and the second is the simplest leading to the same action as full information. The decision maker can choose which one to ask for, depending on the application.

The challenge with eliciting such partitions is assessing whether they contain all relevant information. Only one action is taken, so the decision maker cannot tell if they missed important information regarding untaken actions. If it were instead the case that $p_{a_2} = (0.75, 0.00, 0.05, 0.20)$, then deploying would be the optimal action, but this looks no different under the previously optimal partition $\Pi''$. A mechanism for eliciting partitions must distinguish these two cases, but evaluating the full information partition to do so would defeat the point of simplifying.

\subsection{Our Contribution}

Modeling reward-maximizing agents allows us to construct mechanisms that incentivize simple and decision-relevant partitions, analogous to how proper scoring rules incentivize honest and accurate predictions. Decision-relevance is prioritized, with simplicity encouraged only insofar as no important information is lost. The equilibrium dynamics give a principled reason to believe that AI systems will provide the desired reports, even when decision makers cannot verify whether key information has been omitted. 

We show that a single agent cannot be generally incentivized to optimally partition outcomes, but that two agents can be jointly incentivized to do so in equilibrium. The most straightforward mechanism for this is to elicit a partition of outcomes from the first agent, then ask the other agent for a ``second opinion'' that further splits the partition. Both agents are rewarded for simplicity, but if the refined partition leads to a different action than the original, the second agent is rewarded and the first is punished. The first agent is then incentivized to simplify the outcome space only insofar as it does not change the decision, and in equilibrium the second agent will repeat the same partition.

Along with the simplest set of outcomes where no further information changes the decision, we may also be interested in the simplest set of outcomes leading to the best action. This can be elicited by modifying the above mechanism so that the first agent can provide a ``rebuttal'' that further splits the second agent's partition, canceling the punishment if they are able to change the decision back.

The second opinion mechanism can be adapted to elicit both a simplified partition and a prediction over it. Allowing the second agent to challenge either the first agent's prediction or partition keeps the first agent honest on both fronts. Assigning different roles to two agents illustrates the incentives at play, but the same outputs can be achieved with a single role and a memoryless agent.

These mechanisms can be used as a reward structure in reinforcement learning. As a simple demonstration, we conduct an experiment where models observe MNIST images of digits and choose pixel masks to reveal to a judge, allowing for accurate classification with few pixels. We then introduce a practical application by having language models suggest sets of unit tests for evaluating code. Our mechanism, at either the prompt or training level, allows for a less capable judge to be provided with the minimal information needed to determine which of two code samples is better.

\subsection{Literature Review}
The literature on eliciting predictions over a given set of outcomes is extensive, starting with \citet{Brier1950} and \citet{Good1952}. \citet{gneiting2007strictly} establishes properties that apply to all (strictly) proper scoring rules, which are defined as those that (strictly) incentivize reporting actual beliefs. 

Eliciting conditional predictions for use in making decisions was investigated by \citet{othman2010decision}, who showed that it is impossible to use predictions from a single agent to deterministically take the best action available. Follow-up work \citep{chen2011decision, oesterheld2020decision} found partial workarounds in the single agent case, and \citet{hudson2025joint} showed that multiple agents in a zero-sum competition can instead work. These papers focus on eliciting accurate predictions, rather than decision-relevant ones.

\citet{christiano2021eliciting} introduces the Eliciting Latent Knowledge (ELK) problem, of how to generate honest explanations from an AI that understands some world state. Our mechanism addresses this in the case that humans can understand what the AI deems relevant once their attention has been drawn to it.

\citet{blackwell1951comparison} compares information structures on their ability to lead to better choices on all decision problems, while we instead focus on eliciting the simplest optimal information structure for a particular decision problem. \citet{lei2016rationalizing} and \cite{carter2019what} cover minimal sets of evidence necessary for prioritizing an action, and \citet{li2006towards} earlier discussed coarsening states of Markov Decision Processes so that they left the optimal policy intact. These papers aim to characterize the information set, while we introduce methods for eliciting it from an agent.

The problem of simplifying information for a decision maker with attention costs is modeled in \citet{lipnowski2020attention}, where one party without attention cost summarizes for another party. That paper focuses on the case where both parties share a utility function, while our interest is in specifying a utility function leading to optimal simplification.

The mechanism of rewarding agents for changing the principal's mind appeared in \citet{grace2014how}, which suggested that applying it repeatedly would converge to the truth. Debate between two agents under the same assumption of convergence was explored by \citet{irving2018aisafetydebate} and \citet{browncohen2025avoiding}. \citet{kenton2026debate} and \citet{elasky2026debate} overlap with our focus on applying debate to identify when models anticipate outcomes that humans do not want. Our work differs in that outputs are restricted to be partitions, limiting the opportunity for deceptive arguments and allowing for simplification.

\section{Model}
Let $\mathcal{A}$ be a finite set of actions, and let $\Omega$ be a set of outcomes. A partition $\Pi$ of a set $X$ is a finite collection $Y=\{y_1,\ldots,y_k\}$ of nonempty, pairwise disjoint subsets with $\bigcup_{j=1}^k y_j = X$. Given two partitions of $X$, denoted $\Pi_1$ and $\Pi_2$,  we say $\Pi_2$ is a \textit{coarsening} of $\Pi_1$ (and $\Pi_1$ is a \textit{refinement} of $\Pi_2$) if every element of $\Pi_1$ is wholly contained in some element of $\Pi_2$. A coarsening (resp. refinement) is \textit{strict} if $|\Pi_2|<|\Pi_1|$ (resp. $|\Pi_2|>|\Pi_1|)$.

Let $\mathcal{C}_\Pi$ denote all coarsenings of $\Pi$ and $\mathcal{C}^{>}_\Pi$ strict coarsenings, with $\mathcal{R}_\Pi$ and $\mathcal{R}^{>}_\Pi$ the equivalent for refinements. Let $\mathcal{C}_\Omega$ denote the set of all partitions of $\Omega$.

Given a partition $\Pi$ of $\Omega$, a set of conditional predictions $P \in (\Delta\Pi)^{\mathcal{A}}$ consists of a prediction $p_a \in \Delta\Pi$ for each $a \in \mathcal{A}$. The probability assigned to outcome $\pi \in \Pi$ conditional on action $a$ is denoted $p_{a,\pi}$. For $\Pi_C \in \mathcal{C}_\Pi$, and prediction $P$ on $\Pi$ we let $P_{\Pi_C}$ denote the set of predictions over $\Pi_C$ where $p_{a,\pi_C} = \sum_{\pi \subset \pi_C} p_{a,\pi}$.

We have two agents, denoted $1$ and $2$, who are risk-neutral, and only interested in a reward that can be costlessly provided by the principal. Both agents have beliefs $\mu \in (\Delta\Omega)^{\mathcal{A}}$ regarding the distribution of outcomes, with $\mu_a$ denoting the distribution conditional on action $a$. When agent $i$ reports a partition, it is denoted $\Pi_i$, and when they report a prediction, it is denoted $P^i$. 

The \textit{simplicity} of a partition $\Pi \in \mathcal{C}_{\Omega}$ is measured by a function 
$S:\bigcup_{\Pi \in \mathcal{C}_{\Omega}} (\{\Pi\}\times \{P_\Pi\} ) \rightarrow \mathbb{R}$ that is strictly monotone with coarsening: if $\Pi_2 \in \mathcal{C}_{\Pi_1}$ then $S(\Pi_2, P_{\Pi_2}) \geq S(\Pi_1, P_{\Pi_1})$, with strict inequality when $\Pi_2 \in \mathcal{C}^{>}_{\Pi_1}$. We drop the dependence on $P$ and write $S(\Pi)$ when $P$ is given.

Two examples of simplicity functions are (i) cardinality-based $S_{\mathrm{card}}(\Pi, P_\Pi)=-|\Pi|$, and (ii) entropy-based $S_{\mathrm{ent}}(\Pi, P_\Pi)= \sum_{a\in \mathcal{A}}\sum_{\pi \in \Pi} p^{\Pi}_{a,\pi}\log p^{\Pi}_{a,\pi}$ when all outcomes have non-zero probability. The latter measures the simplicity of a partition when it is applied to a particular prediction, $P$. Without loss of generality, we let $S$ have range $[0,1]$, which can be accomplished by applying the sigmoid function $\sigma(x) = \frac{1}{1+e^{-x}}$.

For generality, we will include a distance metric between actions, $d: \mathcal{A} \times \mathcal{A} \rightarrow \mathbb{R}$, and a threshold $\delta > 0$ for how different actions must be for that difference to be relevant. By default, this can be thought of as the discrete metric, where $d(x,y) = 0$ if $x = y$ and $1$ otherwise, along with $\delta =0.5$.

Finally, we have a decision-making principal, who uses a rule $D:\bigcup_{\Pi \in \mathcal{C}_{\Omega}}(\Delta\Pi)^{\mathcal{A}} \rightarrow \mathcal{A}$ to map conditional predictions to actions. Agents are assumed to know the decision rule, since the principal will not want to conceal it.

The principal may either want agents to provide the simplest partition such that no further refinements would change their choice of action by more than $\delta$, or the simplest partition such that they choose an action within $\delta$ of what they would choose under full information. The set of all partitions meeting the first criterion and second criterion respectively are defined as
\begin{equation*}
    \Pi^* = \arg\max_{\Pi \in \mathcal{R}^S_{\delta}} S(\Pi),\qquad\text{and}\qquad
\Pi^{*+} = \arg\max_{\Pi \in \mathcal{R}^S_{\mathrm{final}}} S(\Pi).
\end{equation*}
where $\mathcal{R}^S_{\delta}= \{\Pi|d(D(P_\Pi),D(P_{\Pi'})) < \delta, \forall\ \Pi' \in \mathcal{R}_\Pi\}$, $\mathcal{R}^S_{final} = \{\Pi|d(D(P_\Pi),D(P_{\Pi_\Omega})) < \delta\}$.

\section{Mechanisms and Results}
\subsection{Simplifying Predictions}
We first analyze the case where an initial expert prediction $P_\Omega$ on a finite partition $\Pi_\Omega$ is given. Both agents know $P_\Omega$, and the principal can access it, but determining $D(P_\Omega)$ is in some way expensive. This corresponds to the case where the agents are independent heads on a neural network with the prediction-making head that produces $P_\Omega$, giving them access to the same information and processing.

The principal cannot in general strictly incentivize a single agent to report a partition in $\Pi^*$ or $\Pi^{*+}$ without being able to fully specify their decision rule.

\begin{theorem}\label{theorem1}
    For any simplicity metric $S$, when $\exists a_1, a_2 \in \mathcal{A}$ such that $d(a_1,a_2) \geq \delta$, there does not exist a partition scoring rule $\mathbb{S}:\bigcup_{\Pi \in \mathcal{C}_{\Omega}} (\{\Pi\}\times \{P_\Pi\} ) \rightarrow \mathbb{R}$ such that either $\argmax_{\Pi \in \mathcal{C}_\Omega}\mathbb{S}(\Pi) \in \Pi^*$ for all $P \in \mathcal{P}$ and decision rule $D$ or $\argmax_{\Pi \in \mathcal{C}_\Omega}\mathbb{S}(\Pi) \in \Pi^{*+}$ for all $P \in \mathcal{P}$ and decision rule $D$
\end{theorem}

Proofs are provided in Appendix A. The intuition here is that the maximal coarsening must receive the highest score among partitions, or it will not get reported when it is in $\Pi^*$. However, if it does receive the highest score, then an agent can report it without the decision maker realizing they are missing information that would cause them to prefer a different action.  

To elicit a partition in $\Pi^*$, consider the following setup, which we call the \textit{second opinion mechanism}. First, agent $1$ provides a partition $\Pi_1 \in \mathcal{C}_{\Pi_\Omega}$. Then, agent $2$ provides a refinement, $\Pi_2 \in \mathcal{R}_{\Pi_1} \cap \mathcal{C}_{\Pi_\Omega}$. Rewards are:
\begin{equation*}
    \begin{cases}
        (2 S(\Pi_1) -  S(\Pi_2), S(\Pi_2) - 2S(\Pi_1)) & \text{if } d(D(P_{\Pi_1}), D(P_{\Pi_2})) < \delta\\
        (2 S(\Pi_1) -  S(\Pi_2) - 4, S(\Pi_2) - 2S(\Pi_1) +4) & \text{otherwise} 
    \end{cases}
\end{equation*}

The value $4$ is chosen so that it is strictly greater than $2 S(\Pi_1) -  S(\Pi_2)$, which follows because $S(\Pi)$ is restricted to $[0,1]$. We make scores zero-sum to prevent potential collusion between agents, but this is not necessary to develop the desired incentives. The principal's actual decision does not affect the agent incentives, so the choice of action when $D(P_{\Pi_1}) \ne D(P_{\Pi_2})$ can be left undetermined.

\begin{theorem}
    In the second opinion mechanism, a subgame perfect equilibrium (SPE) exists, and in any SPE $\Pi_2 = \Pi_1 \in \Pi^*$.
\end{theorem}

The intuition here is that the first agent is rewarded proportionally to the simplicity of their partition, but if there exists a refinement that changes the decision then the second agent will be incentivized to provide it, causing the first agent to lose more than they can gain from simplicity alone. As such, the first agent will optimize their score by reporting the simplest partition with no refinements leading to a change of action, which is the definition of $\Pi^*$.

To elicit a partition in $\Pi^{*+}$, we use a similar setup but give the first agent a chance to respond, which we call the \textit{rebuttal mechanism}. First, agent $1$ provides a partition $\Pi_1 \in \mathcal{C}_{\Pi_\Omega}$. Then agent $2$ provides a refinement $\Pi_2 \in \mathcal{R}_{\Pi_1} \cap \mathcal{C}_{\Pi_\Omega}$. Finally, agent $1$ provides a refinement $\Pi_3 \in \mathcal{R}_{\Pi_2} \cap \mathcal{C}_{\Pi_\Omega}$. Rewards are:
\begin{equation*}
    \begin{cases}
        (4S(\Pi_1) - 2S(\Pi_2) + S(\Pi_3), -4S(\Pi_1) + 2S(\Pi_2) - S(\Pi_3))  & \text{if } d(D(P_{\Pi_3}),D(P_{\Pi_1})) < \delta\\
            &\quad\text{and } d(D(P_{\Pi_2}),D(P_{\Pi_1})) < \delta\\
        (4S(\Pi_1) - 2S(\Pi_2) + S(\Pi_3) + 4, -4S(\Pi_1) + 2S(\Pi_2) - S(\Pi_3) - 4) & \text{if } d(D(P_{\Pi_3}),D(P_{\Pi_1})) < \delta\\
            &\quad\text{and } d(D(P_{\Pi_2}),D(P_{\Pi_1})) \ge \delta\\
        (4S(\Pi_1) - 2S(\Pi_2) + S(\Pi_3) - 8, -4S(\Pi_1) + 2S(\Pi_2) - S(\Pi_3) + 8)    & \text{otherwise}
    \end{cases}
\end{equation*}

\begin{theorem}
    In the rebuttal mechanism, an SPE exists, and in any SPE $\Pi_3 = \Pi_2 = \Pi_1 \in \Pi^{*+}$.
\end{theorem}

In contrast to the second opinion mechanism, the second agent is not incentivized to provide a refinement that changes the decision in the cases where some further refinement from the first agent can change it back. If that occurs, the second agent would not receive the bonus for changing the decision, but would lower their score with a less simple partition and a penalty for being rebutted. 

\subsection{Memoryless Elicitation}
An advantage of using an AI to identify decision-relevant outcomes is that they are memoryless. A single system can assume the roles of both experts, without introducing incentives for collusion. To train an AI for this role, we would like to describe a setup where all agents take the same kind of inputs, produce the same kind of outputs, and are evaluated according to the same reward function.

The necessary incentives are provided through the following mechanism. We start by letting $\Pi_0$ be an arbitrary coarsening. Then, we repeatedly show agent $i$ the partition $\Pi_{i-1}$ and ask it to report a new partition $\Pi_{i}$, which is not restricted to be in $\mathcal{R}_{\Pi_{i-1}}$. The process terminates when $\Pi_i=\Pi_j$ for some $j<i$ with $S(\Pi_j)=S(\Pi_k)$ for $k\in\{j,\ldots,i\}$.

To elicit a partition in $\Pi^*$, we assign agent $i$ the following score, creating the \textit{memoryless second opinion mechanism}:
\begin{equation*}
    S(\Pi_i) - S(\Pi_{i-1}) + \mathbbm{1}_{i-1,i} [2 + S(\Pi_{i-1}) + S(\Pi_{i})]  - \mathbbm{1}_{i,i+1}[(2 + S(\Pi_{i}) + S(\Pi_{i+1}))]
\end{equation*}
where $\mathbbm{1}_{i,j}$ indicates $\Pi_{j} \in \mathcal{R}_{\Pi_{i}}$ and $d(D(P_{\Pi_{i}}),D(P_{\Pi_j})) \geq \delta$.

If an agent's refinement changes the decision from the previous partition, it receives a bonus of $2 + S(\Pi_{i-1}) +  S(\Pi_{i})$, and is therefore incentivized to do so if possible. Even if a subsequent refinement changes the decision back, the agent's score will end up higher by $S(\Pi_{i}) - S(\Pi_{i+1}) > 0$.

\begin{theorem}
    In the memoryless second opinion mechanism, an SPE exists, and in any SPE terminates with report $\Pi_i \in \Pi^*$. 
\end{theorem}

To elicit a partition in $\Pi^{*+}$, we assign agent $i$ the following score, creating the \textit{memoryless rebuttal mechanism}:
\begin{equation*}
    S(\Pi_i) - S(\Pi_{i-1}) + 2 * [\mathbbm{1}_{i-1,i} + \mathbbm{1}_{i-2,i-1,i}' - \mathbbm{1}_{i,i+1} - \mathbbm{1}_{i-1,i,i+1}']
\end{equation*}

where $\mathbbm{1}_{i,j, k}'$ indicates $\Pi_{k} \in \mathcal{R}_{\Pi_{j}}$,  $\Pi_{j} \in \mathcal{R}_{\Pi_{i}}$,  $d(D(P_{\Pi_{i}}),D(P_{\Pi_j})) \geq \delta$, and $d(D(P_{\Pi_{i}}),D(P_{\Pi_k})) < \delta$.

The difference between these mechanisms is that the memoryless second opinion mechanism rewards an additional $S(\Pi_i) + S(\Pi_{i-1})$ for changing the decision, and subtracts an additional $S(\Pi_i) + S(\Pi_{i+1})$ if the following agent changes the decision. If both occur, the agent receives a score of $S(\Pi_{i}) - S(\Pi_{i+1})$, greater than the $0$ they would receive for reporting $\Pi_{i} = \Pi_{i - 1}$. Under the memoryless rebuttal mechanism, an agent who changes the decision and then has it changed gets score $S(\Pi_{i}) - S(\Pi_{i-1}) < 0$. So, agents will anticipate that subsequent agents will change the decision whenever possible under the memoryless second opinion mechanism, but only when it cannot be changed further under the memoryless rebuttal mechanism.

\begin{theorem}
    In the memoryless rebuttal mechanism, an SPE exists, and in any SPE terminates with report $\Pi_i \in \Pi^{*+}$.
\end{theorem}

\subsection{Eliciting Simplified Predictions}

We now relax the assumption that an expert prediction $P_\Omega$ is given. Predictions on $\Omega$ may not be provided, or even providable, due to their complexity. Instead, we would prefer to directly elicit predictions in simplified form.

The argument from the Theorem \ref{theorem1} single-agent impossibility result also applies when both partitions and predictions are elicited. Taking some action based on the maximal coarsening is sometimes but not always optimal, which the principal cannot distinguish. All predictions are the same under maximal coarsening, and so do not affect the result.

\begin{theorem}
    For any simplicity metric $S$, when $\exists a_1, a_2 \in \mathcal{A}$ such that $d(a_1,a_2) \geq \delta$, there does not exist a scoring rule $\mathbb{S}$ such that, for all beliefs $\mu$ and decision rules $D$, the agent's expected score $E_{\omega \sim \mu_a}[\mathbb{S}(\Pi_1, P^1_{\Pi_1}, a, \omega)]$ with $a = D(P^1_{\Pi_1})$ attains its maximum, and does so either only if $\Pi_1 \in \Pi^*$ or only if $\Pi_1 \in \Pi^{*+}$, for $\Pi^*$ and $\Pi^{*+}$ evaluated at $P = \mu$.
\end{theorem}

When eliciting conditional predictions, those conditioned on untaken actions cannot be evaluated for accuracy. \citet{hudson2025joint} showed that when two agents with the same beliefs are jointly evaluated, it is possible for the combination of a joint scoring rule and decision rule to be jointly quasi-strictly proper. This means that in every equilibrium the principal's preferred action, denoted $a^*$, is chosen, agents are strictly incentivized to report truthfully for the chosen action, and weakly incentivized to report truthfully for unchosen actions.

To do this, after taking action $a$ and observing outcome $\omega$, the score for agent $i \ne j$ is given by $\mathcal{S}_i(P^i, P^j, \omega) = s(p^i_a, \omega) - s(p^j_a, \omega)$, where $P^i$ and $P^j$ are the sets of predictions made by agents $i$ and $j$ respectively and $s$ is a strictly proper scoring rule with a finite range. The principal takes the action corresponding to the most preferred prediction made by either agent, even if agents disagree.

Our mechanisms can then be combined with a sequential version of the \citet{hudson2025joint} mechanism to elicit quasi-strictly proper predictions over a partition in $\Pi^*$ or $\Pi^{*+}$.

This requires two conditions, the first of which is that a simplicity maximizing partition always exists, and the simplicity measure cannot depend on predictions for untaken actions as this would violate the weak incentive for honesty on untaken actions. Cardinality satisfies this condition. 

\begin{condition}\label{condition1}
    $S(\Pi_1, P_{\Pi_1}) = S(\Pi_1, P_{\Pi_1}')$ for all $P, P'$. Furthermore, for all $P$, $\exists \Pi \in \mathcal{R}^S_\delta$ such that $S(\Pi, P_\Pi) = \sup_{\Pi' \in \mathcal{R}^S_\delta} S(\Pi', P_{\Pi'})$
\end{condition}

The second condition requires decisions based on iterative strict refinements of a partition to converge within $\delta$.

\begin{condition}\label{condition2}
    For any prediction $P$ and sequence of partitions $\{\Pi_k\}_{k=0}^\infty$ where for all $n$, $\Pi_{n+1} \in \mathcal{R}_{\Pi_n}$, there exists some $N \in \mathrm{N}$ such that for all $n,m \geq N$, $d(D(P_{\Pi_m}), D(P_{\Pi_n}))  < \delta$
\end{condition}

To elicit simplified predictions, the \textit{zero-sum second opinion} mechanism works as follows. First, agent $1$ reports a partition, $\Pi_1$ and an associated prediction $P^1_{\Pi_1}$. Agent $2$ can then either make a prediction $P^2_{\Pi_1}$ over the same partition, or suggest a strict refinement $\Pi_2 \in \mathcal{R}_{\Pi_1}^>$ and make a prediction $P^2_{\Pi_2}$ over it. If agent $2$ provides a partition, agent $1$ then makes a prediction over it, $P^1_{\Pi_2}$. Once both agents have made predictions for the same partition, the principal chooses the action corresponding to the prediction they most prefer across both agents, denoted $\mathcal{D}(P^1, P^2)$.

The scores for the agents are broken down into second opinion and prediction components. The former is the score the agents would receive from the second opinion mechanism, comparing $D(P^1_{\Pi_1})$ with $D(P^2_{\Pi_2})$. The case where the second agent responds with only a prediction is treated as $\Pi_2 = \Pi_1$, with the decision maker not changing their action.

The prediction component of the score for agent $i$ is $\alpha(P^i, P^j) (s(P^i, \omega) - s(P^j, \omega))$, where $s$ is a strictly proper scoring rule and both $P^i$ and $P^j$ are made over the same partition. $\alpha$ scales the score for the second agent, so that the expected loss from dishonesty while the other agent predicts honestly outweighs the benefit of changing the decision maker's action. 

\begin{equation*}
    \alpha(P^i, P^j) = \begin{cases}
        1 & \text{if } D(P^1_{\Pi_2}) = D(P^2_{\Pi_2})\\
        \frac{8}{\min (E_{\omega \sim p^i_{\tilde{a}}} [\Delta_{ij}], E_{\omega \sim p^j_{\tilde{a}}} [\Delta_{ji}])}& \text{otherwise}
    \end{cases}
\end{equation*}
where $\Delta_{ij} = s(p^i_{\tilde{a}}, \omega) - s(p^j_{\tilde{a}}, \omega)$ and $\tilde{a} = \mathcal{D}(P^1_{\Pi_2}, P^2_{\Pi_2})$.

Under these conditions, this mechanism is quasi-strictly proper and elicits partitions in $\Pi^*$.
\begin{theorem}
    For the zero-sum second opinion mechanism under Conditions \ref{condition1} and \ref{condition2}, a Nash equilibrium (NE) exists, and in any NE agent $1$ reports some $\Pi_1 \in \Pi^*$, agent $2$ responds with a prediction, $\mathcal{D}(P^1_{\Pi_1}, P^2_{\Pi_1}) = a^*_{\Pi_1}$, and both agents are strictly incentivized to predict honestly for $a^*_{\Pi_1}$ and weakly incentivized to predict honestly for all actions. 
\end{theorem}

Here, it was necessary to relax the solution concept from SPEs to NEs, as SPEs require well-defined behavior even out of equilibrium. When agent $2$ misreports the probability for the chosen action, the optimal response from agent $1$ can be to report as closely as possible in the direction of $\mu$, and so the best response function may not be well-defined. 

\section{Experiments}
Applications of decision-relevant predictions can be divided into two main categories of interest. The first is augmenting human decision making with AI, by providing important information in an easy-to-process form. The second is auditing AI decision making, by showing which outcomes matter to them when making a choice. 

We run two sets of experiments, testing judgment augmentation and oversight of AIs. In the first, a model reveals a minimal set of pixels to help a judge classify MNIST images. For the second, faulty code is evaluated using tests proposed by a model. In all experiments, the same model is used for all experts, and so all experts have the same information by construction. Full implementation details for each experiment are available in Appendix B.

\subsection{MNIST Classification}
Our first experiment is intended as a simple and easily replicable demonstration of the mechanism. Following \cite{irving2018aisafetydebate}, we use the classic MNIST dataset as a playground to test our approach. We first train a judge model to classify the images, including in cases where some or many pixels are masked. We then see whether we can train a model to provide the judge with the smallest possible set of information needed to classify the image correctly. The choice of pixels to reveal stands in as an analogy for more complex arguments in other settings.

For this experiment, the partition of the outcome space is the mask revealed to the judge model, with the full information case $\Pi_\Omega$ being the entire image. These partitions group together all possible images that have the same value for the revealed pixels, so revealing fewer pixels gives a coarser partition. The probability of such a set of images is $1$ if the values of those pixels match the given image, and 0 otherwise.  

The simplicity function we use, $S(\Pi) = \frac{1}{1 + e^{\frac{|\Pi| - 84}{48}}}$, is a sigmoid of the number of pixels, centered at $84$ with temperature $48$ to provide the most training signal.  The decision rule is an application of the judge neural network with parameters $\theta$ to the masked image $I$, $D(P_\Pi) = f_\theta(\Pi(I))$, and the action chosen is the resulting distribution over digits. Total variation distance compares actions, $d(p,q) = \frac{1}{2}\sum_{i=0}^9|p_i - q_i|$, with a threshold of $\delta = 0.25$.

The model we train to provide the decision-relevant predictions takes one or two MNIST images and a mask generated by a previous agent as input, then outputs a new mask. A random mask is given as input if this is the first agent in the sequence. For training we use Group Relative Policy Optimization Done Right (Dr. GRPO, \citealp{liu2025understanding}), with the reward calculated according to the memoryless second opinion mechanism. Details, including a robustness analysis, are provided in Appendix~\ref{appendix:mnist}.

\begin{figure}[t]
\includegraphics[width=\textwidth]{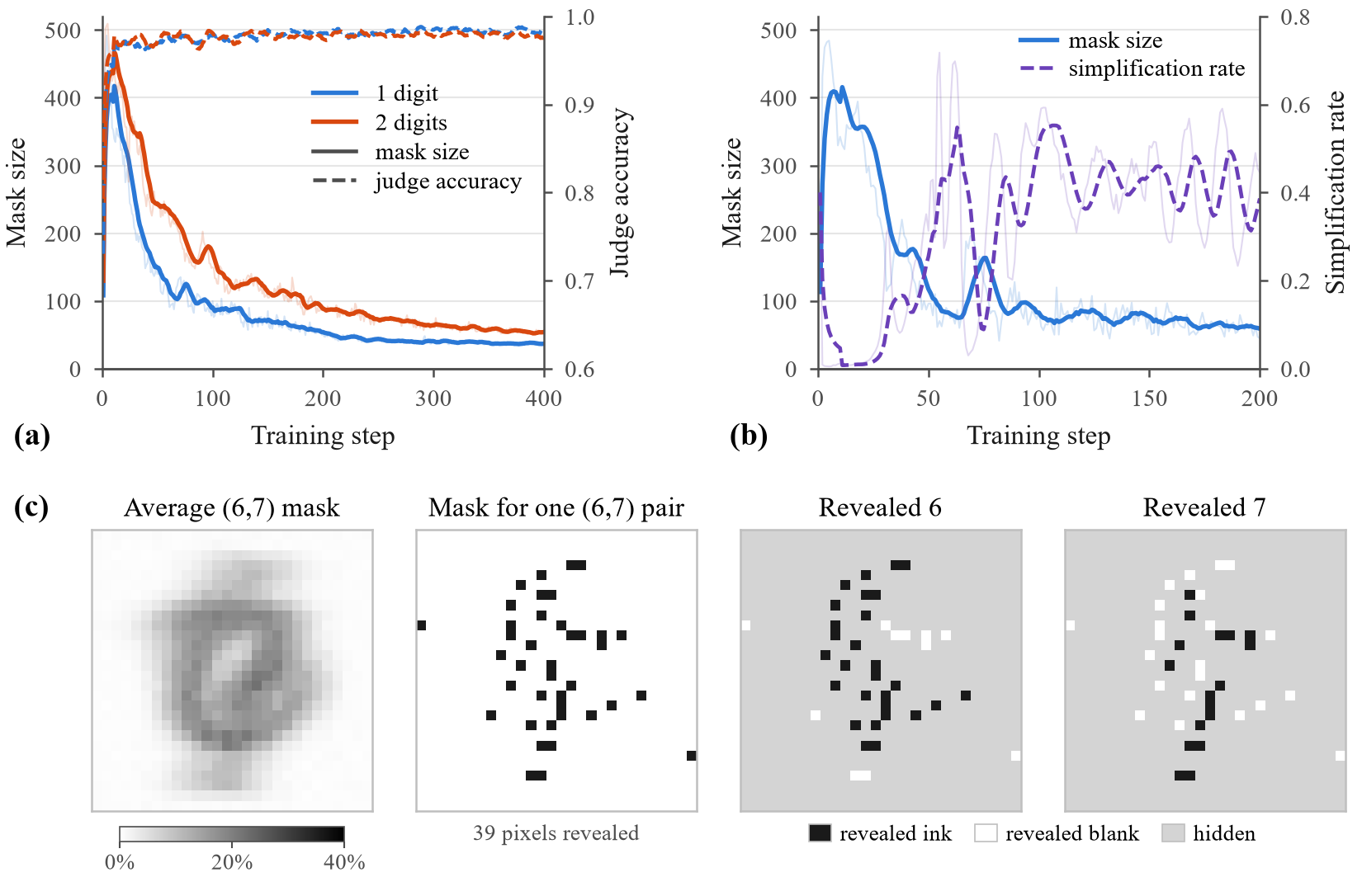}
\caption{MNIST (a) partition simplicity (mask size in pixels) and judge accuracy on masked images over training for one and two digits at once; averaged over four seeds. (b) partition simplicity and rate of simplifying the previous partition for one digit at once; one seed. (c) examples of masks and revealed pixels for a selected two-digit pair.
}
\label{fig:mnist-figure}
\end{figure}

The results can be seen in Figure \ref{fig:mnist-figure}. The top left shows the training curves, where the model learns to provide much smaller masks with no drop in accurate classification. After 400 steps, the average mask size on a held-out dataset was 34.9 for the one image model, and 54.3 for the two image model, with both achieving accuracy rates above 98.5\%.

The top right provides an illustrative example of training dynamics induced by the mechanism. The agent first learns to provide all relevant information, increasing in mask size, then starts to simplify once the judge is reliably answering correctly. In this run, the model briefly overupdates, leading to important information being lost. The mechanism corrects this by then rewarding refinements that change the judge's action towards the full information answer. After a brief spike in mask size, the agent again learns to simplify, but this time cutting only extraneous details.

The bottom row of the figure gives examples of masks for two digits, demonstrating how their features can be combined in under 40 pixels, A single digit typically requires at least 30 pixels to identify.

\subsection{Proposing Tests for Code}

Our second set of experiments tests the ability of LLMs to help a judge decide between two solutions to a coding problem by providing relevant tests. This task is motivated by the reinforcement learning from human feedback (RLHF) process, where a reward model is trained based on pairwise comparisons between generated text. Determining what information is relevant gives insight into what the decision maker values.

We provide an LLM with a coding problem description and two different solutions, then ask for a set of test cases distinguishing the solutions. These sets partition the space of possible code into groups determined by the test outcomes, similar to AlphaCode's approach \citep{Li2022competition}, so adding a test refines the partition. A judge LLM uses the tests to decide which solution is better, assessing the severity of issues when both solutions have bugs.

For code samples, we use solutions to HumanEval problems from the EvalPlus dataset \citep{liu2023codegeneratedchatgptreally}. HumanEval gives 164 short Python problems, to which EvalPlus adds additional tests and candidate solutions from different LLMs. We filter down to solutions written by the Code Llama family of models that pass at least 50\% of provided tests, so that all code shares similar stylistic features and looks plausibly correct. This gives us 2,507 examples of buggy code that fails at least one test and 1,189 examples of correct code that passes all tests. 

Although any set of unit tests could be a partition, we set a maximum for practicality. This does not meaningfully affect results given that the code being compared is fairly simple. We cannot check the full information decision where the judge observes all possible unit tests, but can reasonably assume it matches decisions that are robust to being changed. 

The simplicity metric we use linearly decreases in test count, $S(\Pi) = 1 - \frac{|\Pi|}{|\text{max tests}|}$. The decision rule is an application of the judge neural network with parameters $\theta$ to a prompt containing the set of tests $\Pi$, $D(P_\Pi) = f_\theta(\text{prompt}(\Pi))$, and the action is the resulting distribution over solutions.

\begin{figure}[t]
\centering
\includegraphics[width=\textwidth]{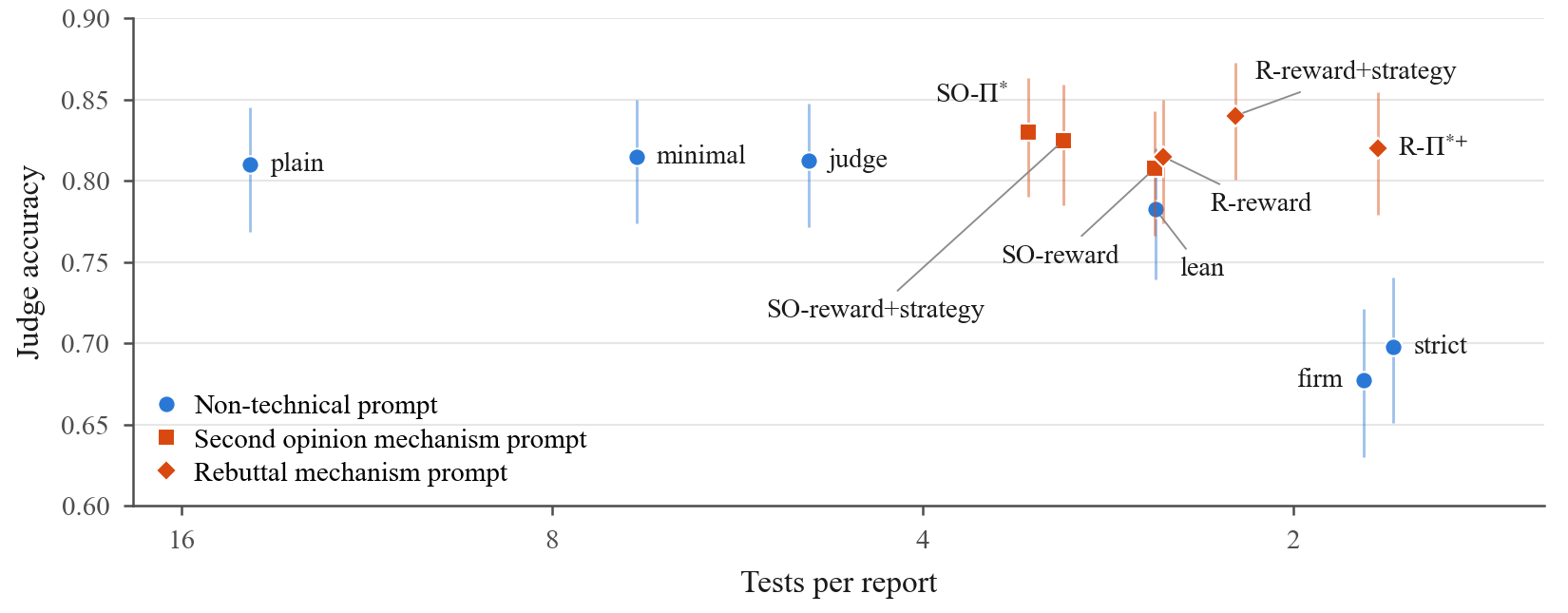}
\caption{
Average number of tests elicited plotted against their effectiveness at leading a judge to the correct decision, for twelve variations on prompts. Prompts were used on Claude Opus 5, with Claude Sonnet 5 judging.
}
\label{fig:drp-code-cascade-claude}
\end{figure}

\paragraph{Prompt-Based Elicitation.} We first evaluate whether we can prompt language models to follow the second opinion and rebuttal mechanisms. We prompt a suite of frontier models with a problem description and a pair of solutions, explain the mechanisms to them, and ask them to provide a set of test cases that will maximize their score. Only the first turn of the game is actually run. We also test describing $\Pi^*$ or $\Pi^{*+}$, and asking for a set of tests in that category. We then compare these to several baseline prompts eliciting tests, varying the intensity with which simplicity and relevance are requested. For the full prompts and extended results see Appendix~\ref{appendix:llm-prompting}. 

In this experiment, the judge is not shown the code itself and must make a decision based on only the tests, representing a situation where the principal is not capable of processing all information themselves and so must rely on well-elicited relevant expert advice.

Figure~\ref{fig:drp-code-cascade-claude} compares twelve prompts for eliciting sets of tests from Claude Opus 5, aimed at a Claude Sonnet 5 judge. Results for other models are included in Appendix~\ref{appendix:llm-prompting}. Simply asking for tests results in nearly 14 provided, on average. Requesting simplicity and relevance with sufficient zeal can reduce the average tests provided to just under 2, but at the cost of usefulness to the judge. 

Prompt variants instructing the agent to behave as if it is optimizing the second opinion mechanism or rebuttal mechanism dominate the non-technical prompts, leading to higher accuracy while also being simpler. It may still be possible to achieve similar performance with traditional prompts, but these results suggest our mechanisms are a straightforward method for improvement.

\begin{figure}[t]
\includegraphics[width=\textwidth]{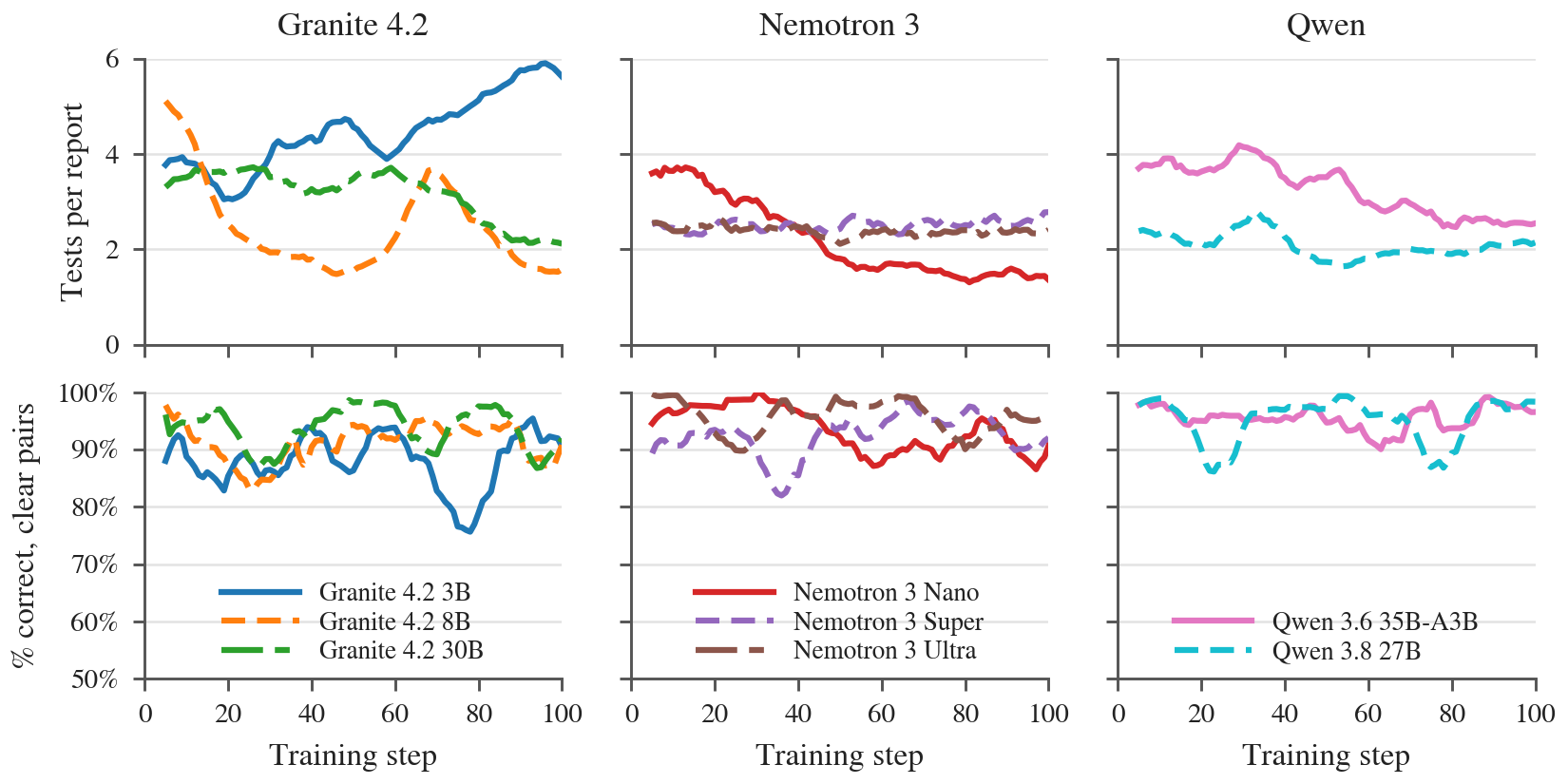}
\caption{(Top) partition simplicity (number of unit tests) over training; (Bottom) judge correctness (choosing correct solutions where possible, and ones with higher pass rate on held out tests when the difference in pass rate is large) over training. Training curves start after five rounds, due to a high rate of initial malformed outputs for some models.
}
\label{fig:training-code-figure}
\end{figure} 

\paragraph{Training-Based Elicitation.} Next, we test using the rewards from the memoryless second opinion mechanism to train open-source LLMs. For each coding task, we draw eight pairs of solutions where exactly one passes all tests, and eight pairs where neither passes all tests. Starting with a random selection of tests from the dataset, we then run the mechanism on each pair, and train on the assigned rewards using Dr. GRPO.

For this experiment, the judge is the same model as the agent, restricted to a single token of output providing logits for each label. The distance metric $d$ measures whether the logits shift by at least $0.5$. Full details are provided in Appendix~\ref{appendix:training-code}.

We train eight models, from three families of open source models: Granite 4.2 (3B, 8B, and 30B), Nemotron 3 (30B-A3B, 120B-A12B, and 550B-A55B), and Qwen (3.6 35B-A3B and 3.8 27B). Results can be seen in Figure \ref{fig:training-code-figure}.

The number of tests per partition drops over training for six of the eight models, increasing slightly for Nemotron 3 120B-A12B and significantly for Granite 4.2 3B. The drop in tests is large for Granite 4.2 8B and 30B, along with Nemotron 3 30B-A3B and Qwen 3.6 35B-A3B, while it is small for Nemotron 3 550B-A55B and Qwen 3.8 27B. Performance was close to flat for all models, with only Nemotron 3 30B-A3B significantly dropping in accuracy, suggesting that simplicity largely did not come at the cost of cutting relevant information. A small drop in accuracy in the training curves is expected, as the model learns the intended behavior of switching decisions away from the correct answer when the previous partition is not in $\Pi^*$.

This pattern of results suggests that the largest, most capable models did not benefit from training. As the previous section demonstrated, frontier models were capable of providing the desired partitions based on only a prompt. On the opposite end, Granite 4.2 3B was too small to learn the mechanism correctly given the difficulty of the task. It was the models in the middle of the size range that benefited most from training, learning to provide simpler sets of tests. We expect that this is not a feature of absolute size and capability, but rather relative to the difficulty of the task.

\section{Extensions}
Secondary to the elicitation of all decision-relevant information, an important motivation for this research direction is reducing the cognitive costs on human decision makers. Our model of decision rules is general, but we expand on explicit models of cognitive costs in Appendix C.

The theoretical results presented here have many applications, and one we are particularly excited by is eliciting decision-relevant predictions for multiple decision makers, then combining the partitions to cut to the heart of disagreements between them. This could be extremely useful in AI alignment, helping identify cases where a model cares about different outcomes than intended. Ambitiously, we would like to see an actor-critic setup for general AI, where the critic not only predicts reward, but also the relevant features of the environment that will lead to high reward. Applications like this hint at the enormous value that would be unlocked if we are able to get principled reads on what powerful AI systems value and believe.

\pagebreak
\subsection*{Acknowledgments}
We are grateful to the UK AI Security Institute for providing funding that supported this work.

\subsection*{AI use statement}
In this work, we used generative AI tools to generate a synthetic dataset and to aid or polish writing. Specifically, we used an LLM judge to label data for the prompt-based elicitation experiments. AI was further used in writing the code for experiments, and to check human-written proofs. We also used AI tools to review our writing and suggest better phrasing to improve readability.

We have \textbf{not} used generative AI tools to: help develop theoretical models or conceptual frameworks, formulate mathematical claims, provide critical ingredients for proving mathematical claims, propose or refine hypotheses, design or provide feedback on research  methodology or experiments, clean and reformat dataset, support qualitative and thematic data analysis, interpret results; and assisting with translation is not applicable to this work.

We have reviewed all AI-assisted work. LLM-generated code was reviewed and tested for correctness by at least 1 author. All text rephrases suggested by AI tools were manually verified for correctness.

We take responsibility for the final content of this work, including text, claims or artifacts produced with the aid of generative AI.

\bibliographystyle{plainnat}
\bibliography{Bibliography-File.bib}

\pagebreak
\appendix
\section{Appendix A: Proofs}
\setcounter{theorem}{0}

\begin{theorem}
    For any simplicity metric $S$, when $\exists a_1, a_2 \in \mathcal{A}$ such that $d(a_1,a_2) \geq \delta$, there does not exist a partition scoring rule $\mathbb{S}:\bigcup_{\Pi \in \mathcal{C}_{\Omega}} (\{\Pi\}\times \{P_\Pi\} ) \rightarrow \mathbb{R}$ such that either $\argmax_{\Pi \in \mathcal{C}_\Omega}\mathbb{S}(\Pi) \in \Pi^*$ for all $P \in \mathcal{P}$ and decision rule $D$ or $\argmax_{\Pi \in \mathcal{C}_\Omega}\mathbb{S}(\Pi) \in \Pi^{*+}$ for all $P \in \mathcal{P}$ and decision rule $D$
\end{theorem}

\begin{proof}
    Let $\mathbb{S}$ be a partition scoring rule and let $\Pi_0 = \{\Omega\}$ be the maximal coarsening.

    It must be that $\Pi_0 = \argmax_{\Pi \in C_\Omega} \mathbb{S}(\Pi, P_\Pi)$ for all $P \in \mathcal{P}$. If it is not for some $P$, then consider the decision rule $D$ such that $D(P_\Pi) = D(P_{\Pi_0})$ for all $\Pi$. Then, $\Pi^* = \{\Pi_0\}$, so  $\argmax_{\Pi \in \mathcal{C}_\Omega}\mathbb{S}(\Pi, P_\Pi) \not\in \Pi^*$.

    For actions  $a_1, a_2$ such that $d(a_1, a_2) > \delta$, predictions $P^{(1)}$ and $P^{(2)}$ with $p^{(1)}_{a_1} = p^{(2)}_{a_1}$, and partition $\Pi'$, consider a decision rule $D$ such that $D(P^{(1)}_\Pi) = a_1$ for all $\Pi$, $D(P^{(2)}_\Pi) = a_1$ for all $\Pi \ne \Pi'$, and $D(P^{(2)}_{\Pi'}) = a_2$. Then, $\Pi_0 \notin R^S_\delta$ for $P^{(2)}$ but $\Pi_0 \in R^S_\delta$ for $P^{(1)}$. 

    Reporting partition $\Pi_0$ must maximize $\mathbb{S}(\Pi, P^{(1)}_\Pi)$, but cannot maximize $\mathbb{S}(\Pi, P^{(2)}_\Pi)$. If the agent reports $\Pi_0$, then action $a_1$ is chosen for both $P^{(1)}$ and $P^{(2)}$. Since only the chosen action's outcome can be subsequently evaluated, and $p^{(1)}_{a_1} = p^{(2)}_{a_1}$, reporting $\Pi_0$ for $P^{(2)}$ must be optimal, even though it is not in $\Pi^*$. As such, there always exists some $P$ and $D$ such that the optimal partition is not in $\Pi^*$. The same argument also applies to $\Pi^{*+}$.   
\end{proof}

\begin{theorem}
    In the second opinion mechanism, a subgame perfect equilibrium (SPE) exists, and in any SPE $\Pi_2 = \Pi_1 \in \Pi^*$.
\end{theorem}
\begin{proof}
    We prove this with backward induction over the game tree. Given $\Pi_1$, the only way for the second agent to achieve their highest score of $S(\Pi_2) - 2 S(\Pi_1) + 4$ is to report $\Pi_2 \in \mathcal{R}_{\Pi_1} \cap \mathcal{C}_{\Pi_\Omega}$ such that $d(D(P_{\Pi_1}),D(P_{\Pi_2})) \geq \delta$. In any SPE they will do so if that set is non-empty. If there is no such partition, then the second agent will report $\Pi_2= \Pi_1$, which gets them a score of $-  S(\Pi_1)$, higher than they would receive for any other report as $S(\Pi_1) > S(\Pi')$ for all $\Pi' \in \mathcal{R}^{>}_{\Pi_1}$.

    In anticipation of this behavior, the first agent will not report any $\Pi_1 \in \mathcal{C}_{\Pi_\Omega}$ where $ \exists \Pi_2 \in \mathcal{R}_{\Pi_1}$ such that  $d(D(P_{\Pi_1}),D(P_{\Pi_2})) \geq \delta$. Doing so would result in a score below $0$, lower than the minimum score for any partition where no such refinement exists. As reporting any other $\Pi_1$ will give the first agent a score of $S(\Pi_1)$, since no refinement can sufficiently change the decision, the first agent will optimize that score by choosing the simplest partition that meets the constraint. Any $\Pi_1$ that has been so chosen is definitionally in $\Pi^*$.

    Given that the first agent will report some $\Pi_1 \in \Pi^*$,and there will not exist any refinements which would change the action chosen to one a distance greater than $\delta$ from $D(P_{\Pi_1})$, the second agent will report $\Pi_2 = \Pi_1$. This is an SPE, so one always exists, and if either agent does not follow this behavior, it is not a SPE.
\end{proof}

\begin{theorem}
    In the rebuttal mechanism, an SPE exists, and in any SPE $\Pi_3 = \Pi_2 = \Pi_1 \in \Pi^{*+}$.
\end{theorem}

\begin{proof}
    We prove this with backward induction. Starting with $\Pi_3$, the first agent's score is higher by $8$ if $d(D(P_{\Pi_3}),D(P_{\Pi_1})) < \delta$. If $(D(P_{\Pi_2}),D(P_{\Pi_1})) < \delta$, then the agent will report $\Pi_3 = \Pi_2$, as this is the simplest partition available in $\mathcal{R}_{\Pi_2} \cap \mathcal{C}_{\Pi_\Omega}$, and will otherwise choose among the simplest partitions that lead to a sufficiently close decision if any exist.

    Knowing this and given $\Pi_1$, the only way for the second agent to achieve their highest score of $-4S(\Pi_1) + 2S(\Pi_2) - S(\Pi_3) + 8$ is to report $\Pi_2 \in \mathcal{R}_{\Pi_1} \cap \mathcal{C}_{\Pi_\Omega}$ such that $d(D(P_{\Pi_1}),D(P_{\Pi'})) \geq \delta$ for all $\Pi' \in \mathcal{R}_{\Pi_2} \cap \mathcal{C}_{\Pi_\Omega}$. In any SPE they will do so if that set is non-empty, choosing the simplest among them. If there is no such partition, then the second agent will report $\Pi_2 = \Pi_1$, which will get them a score of $-3 S(\Pi_1)$ once it triggers $\Pi_3 = \Pi_2$, higher than they would receive for any other report as $S(\Pi_1) > S(\Pi')$ for all $\Pi' \in \mathcal{R}^{>}_{\Pi_1}$.

    In anticipation of this behavior, the first agent will not report any $\Pi_1 \in \mathcal{C}_{\Pi_\Omega}$ where $ \exists \Pi_2 \in \mathcal{R}_{\Pi_1}$ such that $d(D(P_{\Pi_1}),D(P_{\Pi'})) \geq \delta$ for all $\Pi' \in \mathcal{R}_{\Pi_2} \cap \mathcal{C}_{\Pi_\Omega}$. Doing so would result in a score for them of $4S(\Pi_1) - 2S(\Pi_2) + S(\Pi_3) - 8$ after later reporting $\Pi_3 = \Pi_2$, which is lower than the minimum score of $3 S(\Pi_1)$ for any report where no such refinement exists. The first agent will then optimize that score by choosing the simplest partition that meets the constraint and plan to set $\Pi_3 = \Pi_2$. Any $\Pi_1$ that has been so chosen is definitionally in $\Pi^{*+}$.
    
    Given that the first agent will report some $\Pi_1 \in \Pi^{*+}$, and there will not exist any refinements for which all further refinements  would change the action chosen to one a distance greater than $\delta$ from $D(P_{\Pi_1})$, the second agent will report $\Pi_2 = \Pi_1$, and the first agent will report $\Pi_3 = \Pi_2$. This is an SPE, so one always exists, and if either agent does not follow this behavior, it is not a SPE.
\end{proof}

\begin{theorem}
    In the memoryless second opinion mechanism, an SPE exists, and in any SPE terminates with report $\Pi_i \in \Pi^*$.
\end{theorem}

\begin{proof}

    First, we show that in any SPE, if $\Pi_{i-1} \in \mathcal{R}^S_\delta$, then $\Pi_{i} \in \Pi^*$. Since $\Pi_{i-1} \in \mathcal{R}^S_\delta$, it must be that $\mathbbm{1}_{i-1,i} = 0$. If $\Pi_{i} \in \mathcal{R}^S_\delta$, then agent $i$'s score is definitionally maximized by $\Pi_{i} \in \Pi^*$, which gives a score of at least $0$. 
    
    If $\Pi_i \not\in \mathcal{R}^S_\delta$, then it is a dominant strategy for agent $i+1$ to report some $\Pi_{i+1} \in \mathcal{R}_{\Pi_{i}}$ such that $d(D(P_{\Pi_{i}}),D(P_{\Pi_{i+1}})) \geq \delta$. 
    
    If there exists such a $\Pi_{i+1} \in \mathcal{R}^S_\delta$, this earns a score of $2S(\Pi_{i+1}) + 2$. Reporting either $\Pi_{i+1} \in \mathcal{R}_{\Pi_{i}}$ such that $d(D(P_{\Pi_{i}}),D(P_{\Pi_{i+1}})) < \delta$ or $\Pi_{i+1} \not\in \mathcal{R}_{\Pi_{i}}$ leads to a score of $S(\Pi_{i+1}) - S(\Pi_{i}) - \mathbbm{1}_{i+1,i+2}[(2 +S(\Pi_{i+1}) + S(\Pi_{i+2}))]$, which must be lower.

    If there does not exist such a $\Pi_{i+1} \in \mathcal{R}^S_\delta$, then agent $i+1$'s score for $\Pi_{i+1} \in \mathcal{R}_{\Pi_{i}}$ such that $d(D(P_{\Pi_{i}}),D(P_{\Pi_{i+1}})) \geq \delta$ will be $2S(\Pi_{i+1}) + 2 - \mathbbm{1}_{i+1,i+2}[(2 + S(\Pi_{i+1}) + S(\Pi_{i+2}))] > 0$. Reporting $\Pi_{i+1} \in \mathcal{R}_{\Pi_{i}}$ such that $d(D(P_{\Pi_{i}}),D(P_{\Pi_{i+1}})) < \delta$ leads to a score of $S(\Pi_{i+1}) - S(\Pi_{i})  - \mathbbm{1}_{i,i+1}[(2 + S(\Pi_{i+1}) + S(\Pi_{i+2}))] < 0$.

    Reporting $\Pi_{i+1} \not\in \mathcal{R}_{\Pi_{i}}$  leads to a score of $S(\Pi_{i+1}) - S(\Pi_{i})  - \mathbbm{1}_{i+1,i+2}[(2 + S(\Pi_{i+1}) + S(\Pi_{i+2}))]$. Suppose this is the optimal partition to report. Then, it must be that $S(\Pi_{i+1}) > S(\Pi_{i})$ and $\mathbbm{1}_{i+1,i+2} = 0$, meaning agent $i+2$ does not report $\Pi_{i+2} \in \mathcal{R}_{\Pi_{i+1}}$ such that $d(D(P_{\Pi_{i+1}}),D(P_{\Pi_{i+2}})) \geq \delta$. 

    This cannot be because $\Pi_{i+1} \in \mathcal{R}^S_\delta$. If there exists $\Pi_{i+1} \in\mathcal{R}^S_\delta$ such that $S(\Pi_{i+1}) - S(\Pi_{i})  - \mathbbm{1}_{i+1,i+2}[(2 + S(\Pi_{i+1}) + S(\Pi_{i+2}))] > 0$, then $S(\Pi_{i+1}) > S(\Pi_{i})$, and agent $i$ could have strictly increased their score by reporting $\Pi_{i+1}$ instead of $\Pi_{i}$, contradicting that this is an SPE.

    If $\Pi_{i+1} \not\in \mathcal{R}^S_\delta$ but $d(D(P_{\Pi_{i+1}}),D(P_{\Pi_{i+2}})) < \delta$, agent $i+2$ must be foregoing a score of $2 S(\Pi_{i+2}') + 2 - \mathbbm{1}_{i+2,i+3}[(2 + S(\Pi_{i+2}') + S(\Pi_{i+3}))] > 0$ for $S(\Pi_{i+2}) - S(\Pi_{i+1})  - \mathbbm{1}_{i+2,i+3}[(2 + S(\Pi_{i+2}))]$, and so $S(\Pi_{i+2}) > S(\Pi_{i+1})$. But then agent $i+1$ could strictly increase their score by reporting $\Pi_{i+2}$ instead of $\Pi_{i+1}$, contradicting that this is an SPE.
    
    As such, it must be that if $\Pi_{i-1} \in \mathcal{R}^S_\delta$, then $\Pi_{i} \in \Pi^*$.

    When $\Pi_{i-1} \not\in \mathcal{R}^S_\delta$, then by the same logic as above, in any SPE it must be that either $\Pi_{i} \in \mathcal{R}_{\Pi_{i-1}}$ and $d(D(P_{\Pi_{i-1}}),D(P_{\Pi_{i}})) \geq \delta$ or $i-1 = 0$. Either way, due to the finite size of $\Pi_\Omega$, repeated refinements will eventually lead to some agent $n-1$ reporting $\Pi_{n-1} \in \mathcal{R}^S_\delta$, which is then followed by  $\Pi_{n} \in \Pi^*$. As such, this is the only equilibrium.
    \end{proof}

    \begin{theorem}
    In the memoryless rebuttal mechanism, an SPE exists, and in any SPE terminates with report $\Pi_i \in \Pi^{*+}$.
    \end{theorem}

    \begin{proof}  

    If $\Pi_i \not \in \mathcal{R}^S_{final}$ or $\Pi_{i-1} \in \mathcal{R}^S_{final}$, $\Pi_{i} \in \mathcal{R}^>_{\Pi_{i-1}}$, and $d(D(\Pi_{i-1}),D(\Pi_{i})) \geq \delta$ agent $i+1$ can earn a score of at least $S(\Pi_{\Omega}) - S(\Pi_i) + 2$ by reporting $\Pi_{i+1} = \Pi_\Omega$, and so their optimal response must lead to at least one of $\mathbbm{1}_{i,i+1} = 1$ and $\mathbbm{1}_{i-1,i,i+1}' = 1$. If both conditions hold, agent $i+1$ can earn a score of at least $S(\Pi_{\Omega}) - S(\Pi_i) + 4$ by reporting $\Pi_{i+1} = \Pi_\Omega$, and so their optimal response must lead to both of $\mathbbm{1}_{i,i+1} = 1$ and $\mathbbm{1}_{i-1,i,i+1}' = 1$. 

    As such, in any equilibrium, if one of $\Pi_i \not \in \mathcal{R}^S_{final}$ or $\Pi_{i-1} \in \mathcal{R}^S_{final}$, $\Pi_{i} \in \mathcal{R}^>_{\Pi_{i-1}}$, and $d(D(\Pi_{i-1}),D(\Pi_{i})) \geq \delta$, it must be that $\mathbbm{1}_{i-1,i} = 1$ and $\mathbbm{1}_{i-2,i-1,i}' = 1$. Otherwise, agent $i$ could earn a strictly higher score of $0$ by reporting $\Pi_{i+1} = \Pi_i$, terminating the mechanism. Similarly, it cannot be that both $\Pi_i \not \in \mathcal{R}^S_{final}$ as well as $\Pi_{i-1} \in \mathcal{R}^S_{final}$, $\Pi_{i} \in \mathcal{R}^>_{\Pi_{i-1}}$, and $d(D(\Pi_{i-1}),D(\Pi_{i})) \geq \delta$.

    As such, if $\Pi_{i-1} \in \mathcal{R}^S_{final}$, then $d(D(\Pi_{i-1}),D(\Pi_{i})) \geq \delta$ only if $\Pi_{i} \in \mathcal{R}^S_{final}$, $\Pi_{i} \in \mathcal{R}^>_{\Pi_{i-1}}$, $\Pi_{i-1} \in \mathcal{R}^>_{\Pi_{i-2}}$, $d(D(\Pi_{i-1}),D(\Pi_{i-2})) \geq \delta$ and $d(D(\Pi_{i-2}),D(\Pi_{i})) < \delta$ [and there does not exist $\Pi_{i+1}$ with equivalent properties]. If no such $\Pi_i$ is possible, then the optimal behavior for agent $i$  is reporting $\Pi_{i} \in \Pi^{*+}$, which earns a score of $S(\Pi_{i}) - S(\Pi_{i-1})$ and will be at least zero. This means that the optimal response to a partition in $\mathcal{R}^S_{final}$ will also be in $\mathcal{R}^S_{final}$, and after two responses will always be in $\Pi^{*+}$, and which eventually terminates.

    Finally, we show that this cycle is always started in any equilibrium. From earlier, if $\Pi_i \not \in \mathcal{R}^S_{final}$ then $\Pi_{i-1} \in \mathcal{R}^S_{final}$, $\Pi_{i} \in \mathcal{R}^>_{\Pi_{i-1}}$, and $d(D(\Pi_{i-1}),D(\Pi_{i})) \geq \delta$, as well as $\mathbbm{1}_{i-1,i} = 1$ and $\mathbbm{1}_{i-2,i-1,i}' = 1$. The same also applies to  $\Pi_{i+1}$. However, if it is true for both  $\Pi_i$ and $\Pi_{i+1}$, agent $i$ must be receiving a negative score, and so this cannot be an equilibrium. Therefore in every equilibrium, consecutive partitions reported are never both not in $\mathcal{R}^S_{final}$, and so every equilibrium leads to a report in $\mathcal{R}^S_{final}$ followed by reports in $\Pi^{*+}$ until termination. 

\end{proof}

\begin{theorem}
    For any simplicity metric $S$, when $\exists a_1, a_2 \in \mathcal{A}$ such that $d(a_1,a_2) \geq \delta$, there does not exist a scoring rule $\mathbb{S}$ such that, for all beliefs $\mu$ and decision rules $D$, the agent's expected score $E_{\omega \sim \mu_a}[\mathbb{S}(\Pi_1, P^1_{\Pi_1}, a, \omega)]$ with $a = D(P^1_{\Pi_1})$ attains its maximum, and does so either only if $\Pi_1 \in \Pi^*$ or only if $\Pi_1 \in \Pi^{*+}$, for $\Pi^*$ and $\Pi^{*+}$ evaluated at $P = \mu$.
\end{theorem}

\begin{proof}
    Let $\mathbb{S}$ be a scoring rule and let $\Pi_0 = \{\Omega\}$ be the maximal coarsening. 

    For any belief $\mu$ and action $a$, $\Pi_0$ uniquely maximizes $E_{\omega \sim \mu_a}[\mathbb{S}(\Pi_1, P^1_{\Pi_1}, a, \omega)]$ regardless of the reported $P$. If it is not for some $\mu$ and $a$, then consider the decision rule $D$ such that $D(P^1_{\Pi_1}) = a$ for all $\Pi_1$ and $P^1_{\Pi_1}$. Then, $\Pi^* = \Pi^{*+} = \{\Pi_0\}$, so either expected score does not attain its maximum or does at some $\Pi_1$ not in $\Pi^*$ or $\Pi^{*+}$.

    For actions $a_1, a_2$ such that $d(a_1, a_2) \geq \delta$ and beliefs $\mu$ with $E_{\omega \sim \mu_{a_1}}[\mathbb{S}(\Pi_0, P_{\Pi_0}, a_1, \omega)] \geq E_{\omega \sim \mu_{a_2}}[\mathbb{S}(\Pi_0, P_{\Pi_0}, a_2, \omega)]$, consider a decision rule $D$ such that $D(P_{\Pi_0}) = a_1$ and $D(P^1_{\Pi_1}) = a_2$ for all $\Pi_1 \ne \Pi_0$ and $P^1_{\Pi_1}$. Then, for $P = \mu$, $D(P_\Pi) = a_2$ for all $\Pi \ne \Pi_0$, so $\Pi_0 \notin \mathcal{R}^S_\delta$ and $\Pi_0 \notin \mathcal{R}^S_{\mathrm{final}}$.

    Reporting $\Pi_0$ leads to $a_1$ and reporting any $\Pi_1 \ne \Pi_0$ leads to $a_2$. Only the chosen action's outcome can be evaluated, so the expected score for that is $E_{\omega \sim \mu_{a_2}}[\mathbb{S}(\Pi_1, P^1_{\Pi_1}, a_2, \omega)]$. This is strictly less than $E_{\omega \sim \mu_{a_2}}[\mathbb{S}(\Pi_0, P_{\Pi_0}, a_2, \omega)]$ and therefore strictly less than the expected score for reporting $\Pi_0$. Reporting $\Pi_0$ must be optimal, despite not being in $\Pi^*$ or $\Pi^{*+}$. So, there exists some $\mu$ and $D$ such that the optimal partition is not in $\Pi^*$ or $\Pi^{*+}$.
\end{proof}

\begin{theorem}
    For the zero-sum second opinion mechanism under Conditions \ref{condition1} and \ref{condition2}, a Nash equilibrium (NE) exists, and in any NE agent $1$ reports some $\Pi_1 \in \Pi^*$, agent $2$ responds with a prediction, $\mathcal{D}(P^1_{\Pi_1}, P^2_{\Pi_1}) = a^*_{\Pi_1}$, and both agents are strictly incentivized to predict honestly for $a^*_{\Pi_1}$ and weakly incentivized to predict honestly for all actions. 
\end{theorem}

\begin{proof}
We prove this with backward induction over the game tree. Let $\tilde{a} = D(P^1, P^2)$ denote the action the principal chooses under the agents' reports. Since scores are zero-sum, an agent taking a suboptimal action of the equilibrium path can only increase the payoff to the other agent for reaching that point.

We first show that in any NE both agents predict honestly on the partition over which they jointly predict. By the analysis of \citet{hudson2025joint}, the prediction component of agent $i$'s score is maximized in expectation by $p_{i,\tilde{a}} = \mu_{\tilde{a}}$ whenever agent $j$ predicts honestly, since $s$ is strictly proper. For actions $a \neq \tilde{a}$, outcomes are not observed, so the prediction component does not depend on $p_{i,a}$, and by Condition 1 the simplicity component does not either. The only remaining channel by which a prediction for some action $a$ can affect agent $i$'s score is by changing the action chosen.

The scaling $\alpha_i$ is constructed so that the expected loss from dishonest reporting on $\tilde{a}$ when agent $j$ reports honestly will strictly exceed the maximum possible gain in the second opinion component, so this is never a profitable deviation. If agent $j$ is not reporting honestly, then $\alpha_i$ ensures that the prediction component of agent $i$'s score has an expected value of at least $8$ by reporting honestly. This means that the prediction component of agent $i$'s score has an expected value of less than $-8$, and so cannot be a NE as they can always earn a higher score with honesty.  

Therefore both agents are strictly incentivized to predict honestly for $\tilde{a}$ and weakly incentivized to predict honestly for all actions, and when both do so $D(P^1_\Pi, P^2_\Pi) = D(P_\Pi)$ for every partition $\Pi$. Therefore, after agent 2 reports a strict refinement $\Pi_2 \in \mathcal{R}^{>}_{\Pi_1}$ and prediction $P^2_{\Pi_2}$, agent 1 reports $P^1_{\Pi_2}$ honestly in any NE.

Given $\Pi_1$ and $P^1_{\Pi_1}$, agent 2 chooses between reporting only a prediction $P^2_{\Pi_1}$, which is treated as $\Pi_2 = \Pi_1$, and reporting a strict refinement $\Pi_2 \in \mathcal{R}^{>}_{\Pi_1}$ with a prediction $P^2_{\Pi_2}$. Since both agents predict honestly in any NE, the second opinion component reduces to the original second opinion mechanism applied to the truthful prediction $P$. The only way for agent 2 to achieve a score above $-S(\Pi_1, P^1_{\Pi_1})$ is to report $\Pi_2 \in \mathcal{R}^{>}_{\Pi_1}$ such that $d(D(P_{\Pi_1}), D(P_{\Pi_2})) \geq \delta$, in which case the score is $S(\Pi_2, P^1_{\Pi_2}) - 2 S(\Pi_1, P^1_{\Pi_1}) + 4 > 0$. In any NE agent 2 will do so if such a partition exists, choosing the simplest. If there is no such partition, agent 2 will report only a prediction, receiving a score of $-S(\Pi_1, P^1_{\Pi_1})$, higher than the $S(\Pi_2, P^1_{\Pi_2}) - 2S(\Pi_1, P^1_{\Pi_1}) < - S(\Pi_1, P^1_{\Pi_1})$ they would receive for any strict refinement that does not change the decision.

In anticipation of this behavior, agent 1 will not report any $\Pi_1 \in \mathcal{C}_{\Omega}$ where $\exists \Pi_2 \in \mathcal{R}^{>}_{\Pi_1}$ such that $d(D(P_{\Pi_1}), D(P_{\Pi_2})) \geq \delta$. Doing so would result in a score of $2 S(\Pi_1, P^1_{\Pi_1}) - S(\Pi_2, P^1_{\Pi_2}) - 4 \leq -2 < 0$, lower than the minimum score of $S(\Pi_1, P^1_{\Pi_1}) \geq 0$ for any partition where no such refinement exists. By Condition 2 such partitions are guaranteed to exist, since along any chain of refinements decisions converge, so some sufficiently fine partition lies in $\mathcal{R}^S_\delta$. As reporting any other $\Pi_1$ will give agent 1 a score of $S(\Pi_1, P^1_{\Pi_1})$, they will optimize that score by choosing the simplest partition that meets the constraint. Any $\Pi_1$ that has been so chosen is definitionally in $\Pi^*$.

Given that agent 1 will report some $\Pi_1 \in \Pi^*$, and there will not exist any refinements that change the action chosen to one a distance greater than $\delta$ from $D(P_{\Pi_1})$, agent 2 will respond with only a prediction $P^2_{\Pi_1}$, and the principal will choose $D(P^1_{\Pi_1}, P^2_{\Pi_1}) = a^*_{\Pi_1}$. This is a NE, so one always exists, and if either agent does not follow this behavior, it is not a NE.
\end{proof}

\section{Appendix B: Experiments}

Code for all experiments is available at \url{https://github.com/Principled-Agents/decision-relevant-predictions-code}.

\subsection{MNIST Classification}
\label{appendix:mnist}
\subsubsection{Setup}

\paragraph{Setting.} Revealing a set of pixels is equivalent to partitioning the space of possible images based on the values of those pixels; refining a partition is equivalent to revealing a superset of pixels. (In this setting we abuse the notation and use $\Pi$ to denote the set of pixels revealed.) The decision $D(P_\Pi)$ is the judge's probability vector over digits given the revealed pixels, the distance function $d$ is total variation distance, and $\delta=0.25$.  For games with two images, one mask is applied to both images, the decision is the pair of the judge's probability vectors, $d$ is the sum of the two per-image total variation distances, and the threshold is $\delta\sqrt{2}$. Simplicity is $S(\Pi)=\sigma(-(n-84)/48)$ for $n$ revealed pixels. We centered it at 84 for faster convergence following several exploratory runs.

\paragraph{Judge.} For our purposes we require a robust judge with a human-like ability to correctly guess a digit even if only a few pixels are visible. For that we trained a small CNN on randomly masked images. Training masks are biased towards ``reasonable masks''. We sample the mask size from a uniform distribution in 20\% of the cases and from lognormal distribution with median of 120 pixels in 80\% of the cases. In 50\% of the masks the pixel positions are then sampled with 20:1 bias towards pixels which contain ink, while in the other half they are sampled uniformly.

Simple convolutional networks are vulnerable to adversarial attacks where small deliberate alterations of the input can dramatically influence the prediction. To make the judge robust against such attacks we made two additions: 

1) We used a MIMO (multi-input multi-output, \citealp{havasi2021trainingindependentsubnetworksrobust}) architecture where the judge learns 3 independent subnetworks that approximates an ensemble of 3 identical models trained on different seeds.

2) We trained against a simple adversarial attack. For each image we select a small sample of extra pixels which, when added to the mask, reduce the probability the judge assigns to the correct digit the most. Then we calculate loss on this mask and add it to the total loss. 

\paragraph{Policy.} As the policy to generate masks we train a UNet which plays every agent as per the memoryless second opinion mechanism. It sees only the image and $\Pi_{i-1}$, the previous image mask. It outputs independent per-pixel reveal probabilities from which we sample a new mask $M$. It also has a second head which decides how we build the new partition $\Pi_i$ from $M$. It can choose from 3 actions:

- \emph{repeat}: discard $M$ and set $\Pi_i=\Pi_{i-1}$

- \emph{refine}: $\Pi_i=\Pi_{i-1}\cup M$

- \emph{new}: $\Pi_i=M$. 

This explicit action selection is needed because it is unlikely that a policy would produce $M$ which is a repeat or a refinement of $\Pi_{i-1}$ naturally, because each pixel is sampled individually. In addition, to boost exploration of different actions, we add an entropy bonus on the action selection head to the training objective with coefficient $0.08$. 

\paragraph{Game and reward.} $\Pi_0$ is a random mask of log-uniform size. Games end by the mechanism's termination rule ($\Pi_i=\Pi_j$ for some $j<i$ with $S(\Pi_j)=S(\Pi_k)$ for $k\in\{j,\ldots,i\}$) or after $8$ turns. Agents receive the memoryless second opinion score. The simplicity and bonus terms are computed exactly with judge queries; the penalty term $\mathbbm{1}_{i,i+1}[2+S(\Pi_i)+S(\Pi_{i+1})]$ is estimated by sampling $4$ replies to $\Pi_i$ from the current policy and taking their mean. (If the game ends on the 8\textsuperscript{th} turn, rather than by triggering the usual termination rule, we sample the $\mathbbm{1}_{i,i+1}$ penalty term as if there were a subsequent turn so that ending the game early does not impact the reward structure.)

\paragraph{Results.} In addition to the training curves provided in the main body of the paper, we conducted an evaluation on a held out set of either 50 digits or 50 pairs of digits.

\begin{table}[t]
\centering
\caption{Evaluation of each seed at the end of training on full resolution MNIST.} 

\label{tab:mnist-eval}
\vspace{0.5\baselineskip} 
\begin{tabular}{@{}llrrrrr@{}}
\toprule
 & & \multicolumn{2}{c}{Mask size (px)} & \multicolumn{2}{c}{Judge correct (\%)} & Decision \\
\cmidrule(lr){3-4} \cmidrule(lr){5-6}
Digits & Seed & Mean & Median & Per digit & All digits & match (\%) \\
\midrule
1 & 0 & 32.9 & 32 & 98.5 & -- & 95.0  \\
 & 1 & 31.3 & 29 & 97.8 & -- & 96.0\\
 & 2 & 40.5 & 38.5 & 99.0 & -- & 97.8 \\
 & 3 & 34.9 & 32 & 99.2 & -- & 96.8 \\
\cmidrule(l){2-7}
 & \textit{all} & 34.9 & 33 & 98.6 & -- & 96.4 \\
\midrule
2 & 0 & 62.2 & 58 & 98.2 & 96.5 & 95.5\\
 & 1 & 49.5 & 48 & 99.5 & 99.0 & 95.0 \\
 & 2 & 50.2 & 49 & 97.9 & 95.8 & 94.2\\
 & 3 & 55.4 & 54 & 98.8 & 97.5 & 95.5 \\
\cmidrule(l){2-7}
 & \textit{all} & 54.3 & 52 & 98.6 & 97.2 & 95.1  \\
\bottomrule
\end{tabular}
\end{table}

\paragraph{Robustness.} We test whether the elicited partitions contain substantial omissions or fail to preserve the full information decision. To check omissions, we add up to 32 additional pixels to the mask one at a time, greedily selected to change the decision the most. The full information decision is checked directly. If we cannot change a decision and it matches the full information decision, we call that stable.

From the baseline, we first switch to penalizing a partition if any of its four follow-ups change the action, rather than taking the average penalty across them. We then change the number of follow-ups used to assess this penalty to eight. Then, we try reducing $\delta$ from $0.25$, which stays as the value used to assess stability. Numbers are the mean over $4$ seeds and $400$ masks per seed.

\begin{table}[t]
\centering
\caption{Step-by-step changes to training and their effect on mask size and stability for single images. Each step keeps the changes above it; changes ($\Delta$) are relative to the
previous row, so rows 4b, 4c, and 4d compare against the previous training $\delta$. Stability is always measured at
$\delta = 0.25$.}
\label{tab:stability-breakdown1}
\begin{tabular}{lrrrr}
\toprule
Step & Mask (px) & $\Delta$ & Stable (\%) & $\Delta$ \\
\midrule
Baseline & 31.4 &       & 40.7 &        \\
2. Penalty: max over 4 follow-ups instead of mean                              & 36.3 & +4.9  & 46.3 & +5.6  \\
3. 8 follow-ups & 37.9 & +1.6  & 55.5 & +9.2  \\
4a. Training $\delta = 0.2$                                                  & 43.9 & +6.0  & 64.1 & +8.6  \\
4b. Training $\delta = 0.1$                                                  & 54.5 & +10.6 & 78.4 & +14.3 \\
4c. Training $\delta = 0.05$                                                 & 65.5 & +11.0 & 85.7 & +7.4  \\
4d. Training $\delta = 0.03$                                                 & 71.5 & +6.0 & 90.6 & +4.9 \\
\bottomrule
\end{tabular}
\end{table}

The results for partitions covering two images are similar to the single image case, except the training runs at $\delta = 0.03$ failed to make progress on simplifying masks.

\begin{table}[t]
\centering
\caption{Step-by-step changes to training and their effect on mask size and stability for image pairs. Each step keeps the changes above it; changes ($\Delta$) are relative to the
previous row, so rows 4b, 4c, 4d compare against the previous training $\delta$. 4d is separated, as those training runs did not lead to simpler masks, and is included only for comparison to the previous table. Stability is always measured at
$\delta = 0.25$.}
\label{tab:stability-breakdown2}
\begin{tabular}{lrrrr}
\toprule
Step & Mask (px) & $\Delta$ & Stable (\%) & $\Delta$ \\
\midrule
Baseline & 46.6  &        & 17.9 &       \\
2. Penalty: max over 4 follow-ups instead of mean                              & 51.5  & +4.9   & 29.6 & +11.7\\
3. 8 follow-ups & 57.6 & +6.1 & 43.2 & +13.6  \\
4a. Training $\delta = 0.2$                                                  & 62.7  & +5.1   & 52.2 & +9.0  \\
4b. Training $\delta = 0.1$                                                  & 99.4  & +36.7  & 78.4 & +26.2 \\
4c. Training $\delta = 0.05$                                                 & 125.9 & +26.5  & 83.5 & +5.1 \\
\midrule
4d. Training $\delta = 0.03$                                      & 256.6 & +130.7 & 88.2 & +4.7  \\
\bottomrule
\end{tabular}
\end{table}

\subsubsection{Ablation results}
\label{appendix:ablation-mnist}

We also ran several ablations on this setup, to check if there were methods to make training more efficient. We ran them on images downsampled to 14x14, leading to different optimal pixel counts.

The entropy bonus is essential to prevent the model from collapsing early on to only using the ``refine'' action. 

Varying $\delta$ moves the tradeoff between accuracy and simplicity.

Correctly defining the simplicity metric is also important: when we use a linear simplicity function ($S(\Pi) = 1 - \frac{n}{196}$, where $n$ is the number of pixels in the mask) the training does not result in small mask sizes. Instead our runs use a sigmoidal simplicity function $S(\Pi) = \sigma\!\left(-\frac{n - \mu}{T}\right)$ with fixed $\mu = 21$, $T = 12$ at $14\times14$ ($\mu = 84$, $T = 48$ at full resolution). We also tested an adaptive $\mu$ which is adjusted during training based on a rolling average of the agent's reported mask size, which also performs well.

\begin{table}[h]
\centering
\caption{Ablations of the MNIST setup, one setting changed per row from the baseline. $p(\text{action})$ are the rates at which the policy chooses different actions late in training. For definition of the ``$\mathcal{R}^S_{\delta}$ candidates rate'' metric see Appendix~\ref{appendix:ablation-mnist}.}
\label{tab:mnist-ablations}
\vspace{0.5\baselineskip}
{\renewcommand{\arraystretch}{1.15}\setlength{\tabcolsep}{3pt}%
\begin{tabular}{lcccccc}
\toprule
Ablation & \shortstack[c]{Average\\mask size\\in pixels\strut} & \shortstack[c]{Correct\\prediction on\\masked image\strut} & $p$(repeat) & $p$(refine) & $p$(new) & \shortstack[c]{$\mathcal{R}^S_\delta$\\candidates\\rate\strut} \\
\midrule
baseline & 15.1 $\pm$ 1.5 & 0.945 $\pm$ 0.023 & 0.16 & 0.46 & 0.38 & 0.775 $\pm$ 0.041 \\
no entropy bonus & 52.5 $\pm$ 6.3 & 0.955 $\pm$ 0.021 & 0.00 & 1.00 & 0.00 & 0.877 $\pm$ 0.032 \\
$\delta = 0.1$ & 57.8 $\pm$ 6.5 & 0.980 $\pm$ 0.014 & 0.00 & 1.00 & 0.00 & 0.845 $\pm$ 0.035 \\
$\delta = 0.5$ & 9.9 $\pm$ 0.7 & 0.910 $\pm$ 0.028 & 0.16 & 0.41 & 0.42 & 0.767 $\pm$ 0.041 \\
linear simplicity & 56.7 $\pm$ 5.8 & 0.970 $\pm$ 0.017 & 0.00 & 1.00 & 0.00 & 0.958 $\pm$ 0.020 \\
adaptive simplicity & 15.4 $\pm$ 1.5 & 0.945 $\pm$ 0.032 & 0.16 & 0.46 & 0.38 & 0.795 $\pm$ 0.056 \\
no MIMO judge & 15.5 $\pm$ 1.1 & 0.943 $\pm$ 0.023 & 0.16 & 0.46 & 0.38 & 0.588 $\pm$ 0.048 \\
judge, no adv.\ term & 22.6 $\pm$ 3.7 & 0.968 $\pm$ 0.018 & 0.17 & 0.50 & 0.33 & 0.335 $\pm$ 0.046 \\
\bottomrule
\end{tabular}}

\end{table}

Techniques used to improve judge robustness do two things. They make it possible to converge to smaller masks without measurable accuracy loss, and they help to achieve a high $\mathcal{R}^S_\delta$ candidates rate.

\textbf{$\mathcal{R}^S_{\delta}$ candidates rate} is an extra metric we use to assess the mask quality. It is the proportion of masks which are not ruled out as members of $\mathcal{R}^S_{\delta}$, the set of partitions such that no refinement can change the judge's decision. We cannot verify directly that a mask is in $\mathcal{R}^S_{\delta}$ since that would require checking every refinement, so instead we try to rule a reported partition $\Pi$ out. It is ruled out if (a) the judge's decision on $\Pi$ differs from the decision on the full image, $d(D(P_\Pi),D(P_{\Pi_\Omega})) \geq \delta$, as $\Pi_\Omega$ refines every partition; (b) a beam search over refinements of $\Pi$, guided by the gradient of the judge's decision margin with respect to the revealed pixels, finds a refinement $\Pi' \in \mathcal{R}_\Pi$ with $d(D(P_\Pi),D(P_{\Pi'})) \geq \delta$; or (c) the union of $\Pi$ with any other already known mask that induces a different decision on the same image changes the decision. A partition that survives all three checks is a \emph{candidate}. The candidate rate is the fraction of final reports that are candidates; it is an upper bound on the fraction in $\mathcal{R}^S_{\delta}$, since a failed search is not a proof that no decision-changing refinement exists.

\subsection{Proposing Tests for Code}
\label{appendix:llm-prompting}
\subsubsection{Prompt-based elicitation}

\paragraph{Setting.} For samples of code, we use solutions to HumanEval problems from the EvalPlus dataset \citep{liu2023codegeneratedchatgptreally}. We first group together tests that have perfect correlation with each other on HumanEval solutions, then filter tests down to the shortest in each group. For code solutions, we filter down to examples written by the Code Llama family of models that pass at least 50\% of provided tests. We then designate solutions which fail at least one test as buggy.

We organize these buggy solutions into pairs. We then label each pair with a strong model, in this case Fable 5.1 (high effort), identifying which one of the solutions is ``better'' according to a rubric similar to the one we give to the judge. We then filter out those pairs where Fable is uncertain, or where Fable judgment on which solution is better disagrees with which solution passes more hidden tests. This gives us a set of pairs of buggy solutions where one is nonetheless better than the other. 
We use pairs of buggy vs buggy solutions because in the correct vs buggy pairs only one solution fails any tests, so it is trivial for the judge to select a better one.

The simplicity metric we use is the number of tests.

For all models (labeler, judge, and test writer) we run both AB and BA ordering of solution pairs to eliminate the position bias. When labeler disagrees on ordering, we drop the pair, when judge disagrees on ordering, we treat the result as UNCERTAIN. For test writer we just treat two orderings as two datapoints to average over.

\paragraph{Judge.}  We use Claude Sonnet 5 (medium effort) as a judge. It is instructed to consider the test outcomes and the severity of each of them failing. Then it is asked to select which of the solutions, according to provided test results, is better at solving the task. See the full judge prompt in Section~\ref{appendix:judge-prompt}.

\begin{figure}[h]
\centering
\includegraphics[width=\textwidth]{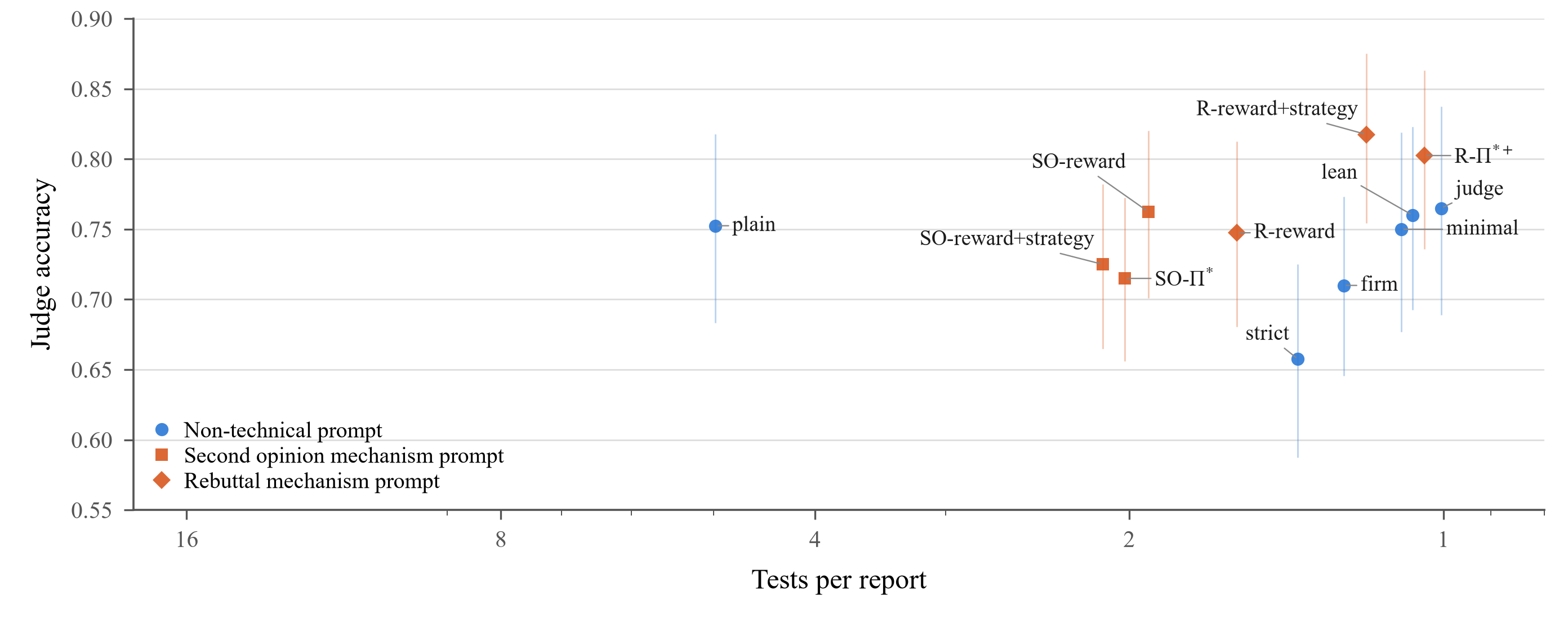}
\caption{
Average number of tests elicited plotted against their effectiveness at leading a judge to the correct decision, for twelve variations on prompts. Prompts were used on ChatGPT Sol 6, with ChatGPT Luna 6 judging.
}
\label{fig:drp-code-cascade-chatgpt}
\end{figure}

\paragraph{Comparison to other models} In addition to Claude Opus 5 and Claude Sonnet 5, we test the same prompts on ChatGPT Sol 6 and ChatGPT Luna 6. Here, the agent is much more prone to providing only a single test, although non-technical prompts result in solo tests that are less helpful. See Figure~\ref{fig:drp-code-cascade-chatgpt}.

\begin{figure}[h]
\centering
\includegraphics[width=\textwidth]{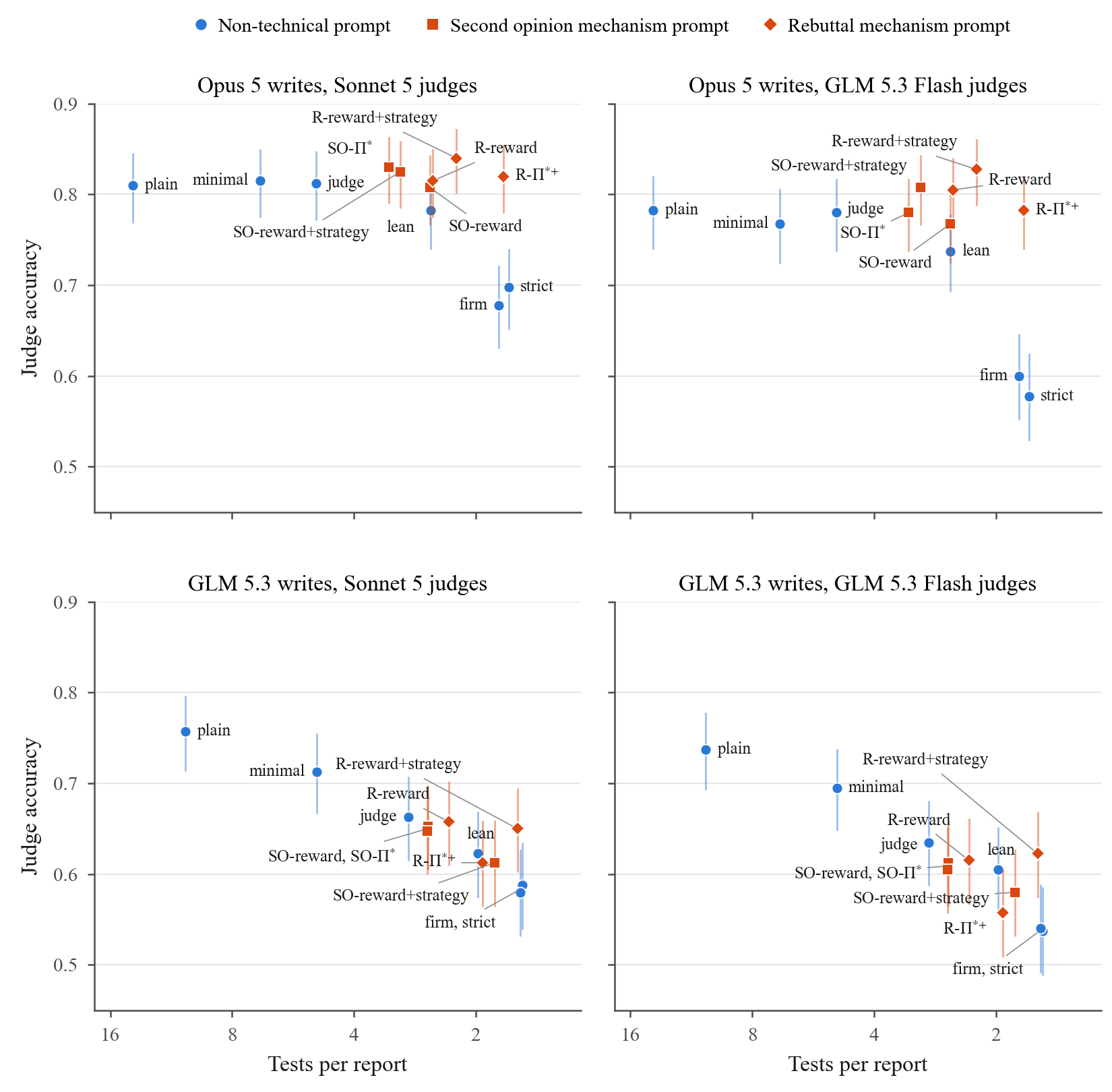}
\caption{
Average number of tests elicited plotted against their effectiveness at leading a judge to the correct decision, for twelve variations on prompts.
}
\label{fig:drp-code-cascade-glm-vs-claude}
\end{figure}

We also test the prompts on GLM 5.3 and GLM 5.3 Flash, then swap judges with the Claude Opus 5 and Claude Sonnet 5 pair to break down the effect by agent and by judge. See Figure~\ref{fig:drp-code-cascade-glm-vs-claude}.

\paragraph{Mechanism-like prompts.} As the unit-test writer agent we use Claude Opus 5 (medium effort). We write three variants of prompts aimed at instilling the same incentives as the mechanisms. In them we explain how model is part of a Second Opinion/Rebuttal game and then either:

1) We describe the reward structure.

2) We describe the reward structure + how a strategy following this reward structure would look like.

3) We describe the definition of the target set ($\mathcal{R}^S_{\delta}$ for the second opinion; $\mathcal{R}^S_{final}$ for the rebuttal) and ask to provide the smallest partition from that set. 

The prompts are listed in Section~\ref{appendix:drp-prompts}.

\paragraph{Control prompts.} As control we use a set of various non-technical prompts which could be reasonably used to elicit the same performance of ``minimal set that gives all required information needed to make a decision''. We vary the wording of the prompts and intensity with which we push for the set to be minimal while still providing the necessary information to the judge. The prompts are listed in Section~\ref{appendix:control-prompts}.

\paragraph{Further Results.} We also tested whether telling the agent that the judge will see the code (and letting the judge see the code) changes the generated set size. In theory, the judge seeing code and being able to reason about it makes some assertions less relevant because the judge can presumably see obvious bugs regardless of whether the expert provides a test that shows them. In the actual experiment on the Agent-Opus/Judge-Sonnet pair we see that it depends on how much you mention in the prompt that the judge can/can't see the code. See results in Table~\ref{tab:judge-sees-code}. Saying to the agent that the judge will not see the code increases the number of asserts, but only if this point is repeatedly mentioned.

\begin{table}[h]
\caption{Whether the judge sees the code, and how the agent is told. Based on the R-reward prompt. (n=100x(2 AB/BA orderings) buggy-vs-buggy pairs; 95\% intervals). Only when told repeatedly that judge can't see the code does it influence the set size the agent provides.}
\label{tab:judge-sees-code}
\vspace{0.5\baselineskip} 
\centering
\begin{tabular}{llcc}
\toprule
judge sees the code & agent told & accuracy (95\% CI) & tests per set (95\% CI) \\
\midrule
no & in passing (default) & 86\% [81, 90] & 2.50 $\pm$ 0.16 \\
no & four times & 88\% [82, 91] & 3.06 $\pm$ 0.17 \\
yes & in passing & 85\% [79, 89] & 2.48 $\pm$ 0.15 \\
yes & four times & 87\% [82, 91] & 2.63 $\pm$ 0.15 \\
\bottomrule
\end{tabular}

\end{table}

\subsubsection{Judge prompt}
\label{appendix:judge-prompt}
The judge's prompt, with the task description and the assertions' outcomes replaced by placeholders; the answer is constrained to a JSON object with the fields the prompt names.
\begingroup\renewcommand{\codingpromptsize}{\scriptsize}\renewcommand{\codingpromptskip}{0.2\baselineskip}
\begin{codingprompt}
[system prompt]\newline
You are a careful reviewer deciding about programs from test evidence alone. You read the outcomes, check each test's expectation against the task description, and never assume a test is right just because it exists; you do not need to solve the task yourself.

[user message]\newline
Two Python solutions, A and B, were written for the task below. You cannot see either solution. A set of test assertions was executed against both; you see each assertion and its outcome on A and on B:\newline
- PASS: the assertion held.\newline
- FAIL: the assertion was false.\newline
- ERROR (type): the assertion raised an exception of that type.\newline
- TIMEOUT: the assertion did not finish within 2 seconds.\newline
- CRASH: the process died while running the assertion.

\#\# Task\newline
\textless{}the task description: the function's signature and docstring\textgreater{}

\#\# Assertions and outcomes\newline
\textless{}the assertions, numbered, each with its outcome on A and on B, one per line:\newline
1. `assert ...`  -\textgreater{}  A: PASS   B: FAIL\newline
...\textgreater{}

\#\# How to judge\newline
- Decide which solution better does what the task asks, from the outcomes above. Your verdict names the better solution: `A` if A does better, `B` if B does better. It never names the worse one: if you find that A is the solution that fails a correct expectation, the verdict is B.\newline
- Start from the assertions whose outcomes differ between A and B: only those can decide the case. An assertion with the same outcome on both cannot favour either solution.\newline
- An outcome counts for or against a solution only insofar as the assertion is a correct expectation for the task. Assertions can be wrong: check each one against the task description yourself. To check an expectation, read the statement and, where needed, work the small input by hand; an input and output taken from the task's own examples are right by definition. You do not need to solve the task in general, design an algorithm for it, or identify where it comes from.\newline
- A PASS also tells you what a solution does on that input; check that behaviour against the task too.\newline
- If the evidence does not distinguish the solutions (no assertions, the same outcomes on both, or outcomes that do not bear on correctness), answer UNCERTAIN. Do not pick a solution on a guess about code you cannot see.\newline
- Keep the reasoning short: the differing outcomes usually settle the case, and if your reasoning is cut off you must decide from what the outcomes show. Before answering, check that the letter you give is the better solution's, not the one you found at fault.

Answer with one JSON object and nothing else. Fill its fields in this order:\newline
- `failures`: one entry per assertion that a solution does not pass (FAIL, ERROR, TIMEOUT, or CRASH): `assertion` (its number), `solution` (`A` or `B`), `expectation\_correct` (whether the assertion's expectation is right for the task), `scope` (what the failure shows about that solution and how much of the task it affects: `misreads\_task` (a misreading of the task or an approach that cannot work), `broad` (a broad class of ordinary inputs handled wrongly), `edge\_case` (a narrow edge case), or `none` (the expectation is wrong, so the failure shows nothing)), and `what\_it\_shows` (a sentence).\newline
- `most\_serious`: for each of `A` and `B`, the most serious flaw its failures show, or that none was shown.\newline
- `weighing`: which solution's flaws matter more for the task as described. The number of assertions on each side is not the measure: one failure that shows a solution misreads the task outweighs several that show edge cases, and an assertion with a wrong expectation counts for nothing.\newline
- `verdict`: `A`, `B`, or `UNCERTAIN`, the better solution, decided from the fields above.\newline
\mbox{}
\end{codingprompt}
\endgroup

\subsubsection{Control prompts}
\label{appendix:control-prompts}
The 6 control prompts. The first, plain, is shown in full, with the task and the two solutions replaced by placeholders; every other prompt variant is given as its difference from the closest earlier variant: the sentences and lines it drops are marked \texttt{-} in red, the ones it adds \texttt{+} in green, in the order they appear.

\begingroup\renewcommand{\codingpromptsize}{\scriptsize}\renewcommand{\codingpromptskip}{0.2\baselineskip}

\noindent\textbf{plain}
\begin{codingprompt}
[system prompt]\newline
You are a meticulous code auditor. You surface flaws in code with test assertions, and you never claim a flaw that is not there.

[user message]\newline
You are comparing two Python solutions, A and B, to the same programming task. Your goal is to surface behaviour that shows which solution does what the task asks better. You do not need to solve the task yourself, trace the programs line by line, or work out where the problem comes from: what matters is finding inputs on which A and B could give different outputs, and knowing the right output for them.

A separate judge decides which solution is better, or answers UNCERTAIN if it cannot tell. The judge sees only the task description, the assertions, and each assertion's outcome on solution A and on solution B (PASS, FAIL, ERROR with the exception type, or TIMEOUT). The judge never sees the code. It answers UNCERTAIN unless the assertions distinguish the solutions, and it checks each assertion against the task description, so an assertion whose expectation is wrong will not help.

Rules for assertions:\newline
- The block holds only plain Python `assert` statements, one per line, inside a single ```python block, and nothing else: no `import`, `def`, `class`, `exec`, assignment, `print`, comment, or prose. A block containing any of these is rejected, and so is one that defines, copies, or reimplements the code under review.\newline
- Each assertion runs on its own against each solution, with the function named in the task already defined (do not define it yourself) and these modules available: math, re, itertools, collections, functools, string, heapq, bisect, operator, fractions, decimal, statistics, copy, json, typing. An assertion that does not finish within 2 seconds is reported as TIMEOUT.\newline
- Prefer assertions whose outcomes differ between the two solutions and whose expectation is clearly right for the task.\newline
- At most 24 assertions in all, including previously proposed assertions; a larger set is rejected.

\#\# The case\newline
\#\#\# Task\newline
\textless{}task description: the function's signature and docstring\textgreater{}

\#\#\# Solution A\newline
```python\newline
\textless{}solution A\textgreater{}\newline
```

\#\#\# Solution B\newline
```python\newline
\textless{}solution B\textgreater{}\newline
```

After your reasoning, output only the code block, with no text before or after it:\newline
```python\newline
assert ...\newline
```\newline
\mbox{}
\end{codingprompt}

\noindent\textbf{minimal}: as \textbf{plain}, with
\begin{codingprompt}
[system prompt]\newline
\textcolor{red!75!black}{- You surface flaws in code with test assertions, and you never claim a flaw that is not there.}\newline
\textcolor{green!45!black}{+ You surface flaws in code with the smallest possible set of test assertions, and you never claim a flaw that is not there.}\newline
[user message]\newline
\textcolor{red!75!black}{- Your goal is to surface behaviour that shows which solution does what the task asks better.}\newline
\textcolor{green!45!black}{+ Your goal is to surface behaviour that shows which solution does what the task asks better, using as little test code as possible.}
\end{codingprompt}

\noindent\textbf{lean}: as \textbf{plain}, with
\begin{codingprompt}
[system prompt]\newline
\textcolor{red!75!black}{- You surface flaws in code with test assertions, and you never claim a flaw that is not there.}\newline
\textcolor{green!45!black}{+ You surface flaws in code with the smallest possible set of test assertions: every assertion you write is one the judge needs, and you never claim a flaw that is not there.}\newline
[user message]\newline
\textcolor{red!75!black}{- Your goal is to surface behaviour that shows which solution does what the task asks better.}\newline
\textcolor{green!45!black}{+ Your goal is to surface behaviour that shows which solution does what the task asks better, with the smallest set of assertions that can show it: nothing the judge does not need to tell the two apart.}\newline
\textcolor{green!45!black}{+ \#\# The set you write}\newline
\textcolor{green!45!black}{+ Write the smallest set of assertions that lets the judge tell which solution is better, and nothing else.}\newline
\textcolor{green!45!black}{+ Every assertion must earn its place:}\newline
\textcolor{green!45!black}{+ - An assertion belongs in the set only if its outcome differs between A and B and that difference bears on which solution does the task better.}\newline
\textcolor{green!45!black}{+ An assertion both solutions pass, or both fail, tells the judge nothing about which is better: leave it out, however informative it looks on its own.}
\end{codingprompt}

\noindent\textbf{firm}: as \textbf{lean}, with
\begin{codingprompt}
[user message]\newline
\textcolor{green!45!black}{+ - Do not add assertions to be thorough, to cover the task, or to show your work.}\newline
\textcolor{green!45!black}{+ The judge does not reward coverage: every extra assertion above the required set is a waste, and every assertion that does not change the judge's decision is dead weight.}\newline
\textcolor{green!45!black}{+ - The only reason to keep more than one assertion is that the judge needs each of them to reach the right decision: for instance, when both solutions are flawed, one assertion showing each solution's flaw, so that the judge can weigh them against each other.}\newline
\textcolor{green!45!black}{+ If a single assertion decides the case, the set is that one assertion.}
\end{codingprompt}

\noindent\textbf{strict}: as \textbf{firm}, with
\begin{codingprompt}
[user message]\newline
\textcolor{red!75!black}{- Every assertion must earn its place:}\newline
\textcolor{green!45!black}{+ Every assertion must earn its place, and the standard is strict:}\newline
\textcolor{green!45!black}{+ - Two assertions that show the same flaw on the same kind of input are one too many: keep the clearest, drop the rest.}\newline
\textcolor{green!45!black}{+ - Before you answer, go through your set once more, assertion by assertion, and remove every one whose removal would leave the judge's decision unchanged.}\newline
\textcolor{green!45!black}{+ What remains is your answer.}
\end{codingprompt}

\noindent\textbf{judge}: as \textbf{minimal}, with
\begin{codingprompt}
[user message]\newline
\textcolor{green!45!black}{+ \#\# What the judge looks for}\newline
\textcolor{green!45!black}{+ The judge starts from the assertions whose outcomes differ between A and B: only those can decide the case, and an assertion with the same outcome on both cannot favour either solution.}\newline
\textcolor{green!45!black}{+ It counts an outcome for or against a solution only insofar as the assertion is a correct expectation for the task: it checks each expectation against the task description itself, working small inputs by hand where needed, so an input and output taken from the task's own examples are right by definition, and an assertion with a wrong expectation helps neither solution.}\newline
\textcolor{green!45!black}{+ A PASS also tells the judge what a solution does on that input, which it checks against the task too.}\newline
\textcolor{green!45!black}{+ If the evidence does not distinguish the solutions, it answers UNCERTAIN, and it never picks a solution on a guess about code it cannot see.}\newline
\textcolor{green!45!black}{+ It keeps its reasoning short: the differing outcomes usually settle the case.}\newline
\textcolor{green!45!black}{+ So write the smallest set of assertions that decides it: assertions on which A and B differ, whose expectation is clearly right for the task, and whose inputs are small enough for the judge to verify by hand.}\newline
\textcolor{green!45!black}{+ An assertion the judge cannot use this way adds nothing.}
\end{codingprompt}
\endgroup

\subsubsection{Mechanism-like prompts}
\label{appendix:drp-prompts}
The 6 mechanism-like prompts: 3 for the Second Opinion and 3 for the Rebuttal.

\begingroup\renewcommand{\codingpromptsize}{\scriptsize}\renewcommand{\codingpromptskip}{0.2\baselineskip}

\noindent\textbf{SO-reward}: as \textbf{minimal}, with
\begin{codingprompt}
[user message]\newline
\textcolor{green!45!black}{+ \#\# Your role}\newline
\textcolor{green!45!black}{+ How you are scored:}\newline
\textcolor{green!45!black}{+ Setup. Two auditors work on this case, one after the other.}\newline
\textcolor{green!45!black}{+ You are the first: you propose a set of assertions.}\newline
\textcolor{green!45!black}{+ A second auditor, a fresh model that sees the task, the solutions, and your set, then adds any further assertions it likes (possibly none); the final set is your set plus its additions.}\newline
\textcolor{green!45!black}{+ The judge makes separate decisions on your set and on the final set, and the scores depend on whether its decision changes between them.}\newline
\textcolor{green!45!black}{+ The decision is A, B, or UNCERTAIN; with no assertions, or with assertions whose outcomes do not distinguish the solutions, it is UNCERTAIN.}\newline
\textcolor{green!45!black}{+ Simplicity. Every assertion set has a simplicity score S.}\newline
\textcolor{green!45!black}{+ S = 1 - n / 24, where n is the number of assertions in the set: the empty set scores 1, each assertion costs 1 / 24 = 0.042, and a set of 24 assertions scores 0.}\newline
\textcolor{green!45!black}{+ Your score is 2 * S(your set) - S(final set), minus 4 if the judge's decision on the final set differs from its decision on your set (UNCERTAIN to A or B, or A to B or B to A, all count).}\newline
\textcolor{green!45!black}{+ The second auditor's score is the negative of yours.}\newline
\textcolor{green!45!black}{+ An answer that does not follow the format scores -10, and your set is then empty.}\newline
\textcolor{green!45!black}{+ Since the final set contains your set, this comes to S(your set), plus what the second auditor's additions cost in simplicity, minus 4 if they change the decision: each of your own assertions costs you 0.042, each assertion the second auditor adds earns you 0.042, and a changed decision costs you 4, more than any set of assertions is worth.}
\end{codingprompt}

\noindent\textbf{SO-reward+strategy}: as \textbf{SO-reward}, with
\begin{codingprompt}
[user message]\newline
\textcolor{green!45!black}{+ Examples:}\newline
\textcolor{green!45!black}{+ - You propose 2 assertions (S = 0.917) that make the decision A, and no further assertion could change it.}\newline
\textcolor{green!45!black}{+ The second auditor adds nothing: 2 * 0.917 - 0.917 = 0.917.}\newline
\textcolor{green!45!black}{+ If it adds 1 assertion anyway, leaving the decision A: 2 * 0.917 - 0.875 = 0.958.}\newline
\textcolor{green!45!black}{+ - You propose 1 assertion (S = 0.958) that makes the decision A, but a further assertion would turn it to B.}\newline
\textcolor{green!45!black}{+ The second auditor adds it: 2 * 0.958 - 0.917 - 4 = -3.000.}\newline
\textcolor{green!45!black}{+ Proposing both assertions yourself, so that the decision is B and nothing changes it: 0.917.}\newline
\textcolor{green!45!black}{+ - You propose the empty set (S = 1.000), so the decision is UNCERTAIN, and the second auditor adds one assertion that makes it A: 2 * 1.000 - 0.958 - 4 = -2.958.}\newline
\textcolor{green!45!black}{+ What follows for you:}\newline
\textcolor{green!45!black}{+ - Propose the smallest set of assertions that leads the judge to the decision the full evidence supports and that no further assertions could change.}\newline
\textcolor{green!45!black}{+ An assertion you leave out that the second auditor can use to change the decision costs you 4.}\newline
\textcolor{green!45!black}{+ - Do not go beyond that set: each further assertion costs 0.042 and earns nothing.}
\end{codingprompt}

\noindent\textbf{SO-target-set}: as \textbf{minimal}, with
\begin{codingprompt}
[user message]\newline
\textcolor{green!45!black}{+ \#\# Your role}\newline
\textcolor{green!45!black}{+ Setup. Two auditors work on this case, one after the other.}\newline
\textcolor{green!45!black}{+ You are the first: you propose a set of assertions.}\newline
\textcolor{green!45!black}{+ A second auditor, a fresh model that sees the task, the solutions, and your set, then adds any further assertions it likes (possibly none); the final set is your set plus its additions.}\newline
\textcolor{green!45!black}{+ The judge makes separate decisions on your set and on the final set.}\newline
\textcolor{green!45!black}{+ The decision is A, B, or UNCERTAIN; with no assertions, or with assertions whose outcomes do not distinguish the solutions, it is UNCERTAIN.}\newline
\textcolor{green!45!black}{+ The set we want from you has three properties, in this order of importance:}\newline
\textcolor{green!45!black}{+ 1. Right: the judge's decision on your set is the decision the whole truth about the two solutions supports, the one the judge would reach if it saw the outcome of every assertion that could be written.}\newline
\textcolor{green!45!black}{+ 2. Settled: nothing the second auditor could add to your set would change the judge's decision on it.}\newline
\textcolor{green!45!black}{+ Whatever assertions are added, the decision on the final set is the decision on yours: your set already carries the evidence that decides the case and leaves no opening for evidence that would overturn it.}\newline
\textcolor{green!45!black}{+ 3. Smallest: among the sets with the first two properties, yours has the fewest assertions.}\newline
\textcolor{green!45!black}{+ An assertion that is not needed for the decision to be right and to stay settled is left out.}\newline
\textcolor{green!45!black}{+ A set that is right but not settled is not what we want, and neither is one that is settled but not right.}\newline
\textcolor{green!45!black}{+ Smallness comes last: never drop an assertion the first two properties need.}
\end{codingprompt}

\noindent\textbf{R-reward}: as \textbf{minimal}, with
\begin{codingprompt}
[user message]\newline
\textcolor{green!45!black}{+ \#\# Your role}\newline
\textcolor{green!45!black}{+ How you are scored:}\newline
\textcolor{green!45!black}{+ Setup. Two auditors work on this case in three rounds.}\newline
\textcolor{green!45!black}{+ Round 1: you propose a set of assertions.}\newline
\textcolor{green!45!black}{+ Round 2: a second auditor, a fresh model that sees the task, the solutions, and your set, adds any further assertions it likes (possibly none).}\newline
\textcolor{green!45!black}{+ Round 3: you see your set and the second auditor's additions and add any further assertions you like (possibly none); if the second auditor added nothing, there is no round 3 and your set is final.}\newline
\textcolor{green!45!black}{+ Each set contains the one before it.}\newline
\textcolor{green!45!black}{+ The judge makes separate decisions on the first set and on the final set, and the scores depend on whether its decision changes between them; its decision on the second set does not enter the scores.}\newline
\textcolor{green!45!black}{+ The decision is A, B, or UNCERTAIN; with no assertions, or with assertions whose outcomes do not distinguish the solutions, it is UNCERTAIN.}\newline
\textcolor{green!45!black}{+ Simplicity. Every assertion set has a simplicity score S.}\newline
\textcolor{green!45!black}{+ S = 1 - n / 24, where n is the number of assertions in the set: the empty set scores 1, each assertion costs 1 / 24 = 0.042, and a set of 24 assertions scores 0.}\newline
\textcolor{green!45!black}{+ Your score is 4 * S(set 1) - 2 * S(set 2) + S(set 3), minus 8 if the judge's decision on the final set (round 3) differs from its decision on your first set (round 1) (UNCERTAIN to A or B, or A to B or B to A, all count).}\newline
\textcolor{green!45!black}{+ The second auditor's score is the negative of yours.}\newline
\textcolor{green!45!black}{+ An answer that does not follow the format scores -10; in round 1 your set is then empty, in round 3 you then add nothing.}\newline
\textcolor{green!45!black}{+ Since each set contains the one before, this comes to 3 * S(set 1), plus 0.042 for each assertion the second auditor adds, minus 0.042 for each assertion you add in round 3, minus 8 if the final decision differs from the decision on your first set.}\newline
\textcolor{green!45!black}{+ So each assertion in your first set costs you 0.125, and the decision is worth 8.}\newline
\textcolor{green!45!black}{+ The second auditor gains only if its change of the decision survives your round 3, so it will add assertions when it can change the decision and you cannot change it back.}
\end{codingprompt}

\noindent\textbf{R-reward+strategy}: as \textbf{R-reward}, with
\begin{codingprompt}
[user message]\newline
\textcolor{green!45!black}{+ Examples:}\newline
\textcolor{green!45!black}{+ - You propose 2 assertions (S = 0.917) that lead to the decision A, which the full evidence supports.}\newline
\textcolor{green!45!black}{+ The second auditor adds nothing and so do you: 3 * 0.917 = 2.750.}\newline
\textcolor{green!45!black}{+ - You propose 1 assertion (S = 0.958) that leads to the decision A, but the full evidence supports B.}\newline
\textcolor{green!45!black}{+ The second auditor adds 1 assertion that turns the decision to B (S = 0.917), and nothing you add in round 3 turns it back: 4 * 0.958 - 2 * 0.917 + 0.917 - 8 = -5.083.}\newline
\textcolor{green!45!black}{+ Proposing both assertions in round 1 instead: 2.750.}\newline
\textcolor{green!45!black}{+ - You propose 1 assertion (S = 0.958) that leads to the decision A, which the full evidence supports.}\newline
\textcolor{green!45!black}{+ The second auditor adds 1 assertion that turns the decision to B anyway, and in round 3 you add 1 assertion that turns it back to A (S = 0.875): 4 * 0.958 - 2 * 0.917 + 0.875 = 2.875.}\newline
\textcolor{green!45!black}{+ Adding nothing in round 3 instead: 4 * 0.958 - 2 * 0.917 + 0.917 - 8 = -5.083.}\newline
\textcolor{green!45!black}{+ What follows for you:}\newline
\textcolor{green!45!black}{+ - Propose the smallest set of assertions that leads the judge to the decision the full evidence supports, even if the second auditor could add assertions that change the judge's decision: in round 3 you can add assertions that change it back.}\newline
\textcolor{green!45!black}{+ A final decision different from the one your set leads to costs 8; each assertion costs 0.125.}\newline
\textcolor{green!45!black}{+ - In round 3, if the second auditor's additions changed the judge's decision and further assertions would restore it, add the smallest such set; otherwise add nothing.}
\end{codingprompt}

\noindent\textbf{R-target-set}: as \textbf{minimal}, with
\begin{codingprompt}
[user message]\newline
\textcolor{green!45!black}{+ \#\# Your role}\newline
\textcolor{green!45!black}{+ Setup. Two auditors work on this case in three rounds.}\newline
\textcolor{green!45!black}{+ Round 1: you propose a set of assertions.}\newline
\textcolor{green!45!black}{+ Round 2: a second auditor, a fresh model that sees the task, the solutions, and your set, adds any further assertions it likes (possibly none).}\newline
\textcolor{green!45!black}{+ Round 3: you see your set and the second auditor's additions and add any further assertions you like (possibly none); if the second auditor added nothing, there is no round 3 and your set is final.}\newline
\textcolor{green!45!black}{+ Each set contains the one before it.}\newline
\textcolor{green!45!black}{+ The judge makes separate decisions on your set and on the final set.}\newline
\textcolor{green!45!black}{+ The decision is A, B, or UNCERTAIN; with no assertions, or with assertions whose outcomes do not distinguish the solutions, it is UNCERTAIN.}\newline
\textcolor{green!45!black}{+ The set we want from you has two properties, in this order of importance:}\newline
\textcolor{green!45!black}{+ 1. Right: the judge's decision on your set is the decision the whole truth about the two solutions supports, the one the judge would reach if it saw the outcome of every assertion that could be written.}\newline
\textcolor{green!45!black}{+ 2. Smallest: among the right sets, yours has the fewest assertions.}\newline
\textcolor{green!45!black}{+ An assertion that is not needed for the decision to be right is left out.}\newline
\textcolor{green!45!black}{+ That is all.}
\end{codingprompt}
\endgroup

\subsubsection{LLM Training}
\label{appendix:training-code}
\paragraph{Setting.} As above, we use solutions to HumanEval problems from the EvalPlus dataset \citep{liu2023codegeneratedchatgptreally} and applied the same filters. We took a maximum of eight correct solutions and twenty-four incorrect solutions per task and generated two sets of solution pairs: correct-vs-buggy and buggy-vs-buggy. This gave us 835 correct-vs-buggy pairs and 836 buggy-vs-buggy pairs before splitting into train and test sets by problem. During training we sampled evenly from the correct-vs-buggy and buggy-vs-buggy pools.

We capped the number of allowable test cases at 8 and used $S(n) = 1 - \frac{n}{8}$ (where $n$ is the number of tests) as the simplicity function. In principle, models could have hacked this scoring function by combining multiple test cases into one statement, but in our experiments simple instructions sufficed to prevent it. Models were prompted with a description of the memoryless second opinion mechanism, the problem description, the two code samples, and the previous turn's list of assertions, and instructed to respond with an action -- ``done'', ``add'', or ``new'' -- and a Python code block with a list of assertions. Like in the MNIST setting, reporting ``done'' ends the game with a repetition of the previous set of assertions, reporting ``add'' combines the listed assertions with the previous set of assertions, and reporting ``new'' creates a new set of assertions with only those listed by the turn agent. The agents were not provided with the outcomes of either their own test cases or the previous set of test cases. Malformed responses, including those exceeding the reasoning budget of 2048 tokens, were given a fixed score of $-4$.

To start each game, we generated an initial set of assertions to represent $\Pi_0$ by sampling between 0 and 8 assertions from the EvalPlus test cases. These had a 25\% chance of being a test that resolved the same on both solutions, a 12.5\% for each solution of favoring them, and a 50\% chance to be otherwise randomly drawn, repeating the process if a test of the desired type was unavailable. 

\paragraph{Training.} We trained rank-16 LoRAs on Granite 4.2 (3B, 8B, and 30B) \citep{ibm_granite_2026} Nemotron 3 (30B-A3B, 120B-A12B, and 550B-A55B) \citep{nemotron_3_2026}, and Qwen (3.6 35B-A3B and 3.8 27B) \citep{qwen_2026} using Dr. GRPO \citep{liu2025understanding}.

\paragraph{Judge.} For each model, the judge was a copy of it with reasoning turned off. This copy was also trained with a rank-16 LoRA, to match the distribution of responses that it would output with reasoning turned on, on a held out set of comparisons. The judge model was prompted with the problem description, the test cases provided by the models, and their outcomes on each of the code samples (assertions were classified as \texttt{PASS}, \texttt{FAIL}, \texttt{TIMEOUT}, \texttt{ERROR} with the error type provided, or \texttt{CRASH} indicating that the process running the assertion died). It was also given access to the code of the two solutions. The judge was prompted to pick the better solution or to report \texttt{UNCERTAIN} if it could not determine which was better.

We sampled the logits of the judge twice, with each order of the solutions reversed in half of them. A report was treated as changing the judge's decision if it shifted the margin by at least $\delta=0.5$.

\paragraph{Policy.} Models were told to start their response by choosing exactly one action:

- \emph{done}: repeat the previous partition and end the game

- \emph{add}: Add additional tests to the existing suite

- \emph{new}: Provide a new set of tests

Models that chose ``add'' or ``new'' then wrote their set of tests.

\paragraph{Game and reward.} Agents receive the memoryless second opinion score. The simplicity and bonus terms are computed exactly with judge queries; the penalty term $\mathbbm{1}_{i,i+1}[2+S(\Pi_i)+S(\Pi_{i+1})]$ is estimated by sampling $4$ replies to $\Pi_i$ from the current policy and taking their mean. If the game ends on the 5\textsuperscript{th} turn, rather than the usual termination rule, we sample the $\mathbbm{1}_{i,i+1}$ penalty term as if there were a subsequent turn so ending the game early does not impact the reward structure.

\paragraph{Results} In addition to the training curves provided in the main body of the paper, we conducted an evaluation on a held out set of 33 questions, over which we had 186 correct vs. buggy and 182 buggy vs. buggy pairs. Clear pairs are comparisons where one is correct or passes at least 20\% more held out tests than the other. The tests per report and \% correct are measured only on well-formed responses, and only for the first partition provided. For \% correct, we drop a small number of comparisons where the first partition changes the decision away from the correct one, as this is the intended behavior.

\begin{table}[t]
\centering
\small
\begin{tabular}{lrrrrrrrrr}
\toprule
 & & \multicolumn{3}{c}{Tests per report} & \multicolumn{3}{c}{\% correct, clear pairs} & \multicolumn{2}{c}{\% malformed} \\
\cmidrule(lr){3-5} \cmidrule(lr){6-8} \cmidrule(lr){9-10}
Model & Dropped & Start & End & Change & Start & End & Change & Start & End \\
\midrule
Granite 4.2 3B & 22 & 3.98 & 4.98 & $+1.00^{***}$ & 85.6 & 77.7 & $-7.9$ & 2.9 & 1.2 \\
Granite 4.2 8B & 14 & 5.16 & 2.48 & $-2.68^{***}$ & 94.8 & 88.6 & $-6.2$ & 13.8 & 0.0 \\
Granite 4.2 30B & 19 & 4.02 & 3.32 & $-0.70^{***}$ & 91.7 & 90.7 & $-1.0$ & 3.2 & 3.4 \\
\midrule
Nemotron 3 Nano & 6 & 4.08 & 1.80 & $-2.28^{***}$ & 94.0 & 85.3 & $-8.7^{**}$ & 94.8 & 0.0 \\
Nemotron 3 Super & 22 & 3.39 & 3.58 & $+0.19$ & 93.1 & 94.6 & $+1.5$ & 3.2 & 0.3 \\
Nemotron 3 Ultra & 5 & 3.23 & 3.07 & $-0.16$ & 95.4 & 93.6 & $-1.8$ & 37.5 & 0.6 \\
\midrule
Qwen 3.6 35B-A3B & 7 & 4.12 & 2.26 & $-1.86^{***}$ & 94.0 & 89.9 & $-4.1$ & 93.9 & 1.1 \\
Qwen 3.8 27B & 7 & 3.47 & 2.63 & $-0.84^{***}$ & 94.5 & 92.6 & $-1.9$ & 47.4 & 1.4 \\
\bottomrule
\end{tabular}
\caption{Held-out evaluation of each agent before and after training. Each pair of held-out solutions is played once by each agent, with significance judged by a two-sided paired permutation test swapping each pair's two games, then Holm-corrected over all tests in the table. $^{*}p<0.05$, $^{**}p<0.01$, $^{***}p<0.001$ }
\label{tab:coding-eval}
\end{table}

\section{Appendix C: Cognitive Costs}
\label{appendix:cognitive-costs}

A secondary but important benefit of eliciting decision-relevant predictions is reducing the principal's cognitive costs. Such costs can be modeled explicitly.

\subsection{Principal Cognitive Costs}

Consider a principal with separable utility $u(\omega,\Pi)=v(\omega)-c(\Pi)$, where cognitive cost $c$ strictly increases with refinement. Let $S(\Pi)=1-c(\Pi)/c(\Pi_\Omega)$.

Cognitive costs are determined before the principal chooses an action. Let 
\[
    a_\Pi=D(P_\Pi)\in\argmax_{a\in\mathcal A} E_{\omega \sim P_{a,\Pi}}[v(\omega)].
\]

$E_{\omega\sim P_{a,\Pi}}[v(\omega)]$ is the principal’s subjective expected utility, based on reported probabilities for partition cells and the principal’s conditional beliefs about outcomes within cells. We assume $E_{\omega \sim P_{a,\Pi_\Omega}}[v(\omega)]=E_{\omega\sim p_a}[v(\omega)]$. Let $a^*=a_{\Pi_\Omega}$ and $W(\Pi)=E_{\omega\sim p_{a_\Pi}}[v(\omega)]-c(\Pi)$. Additionally, let $d$ be the discrete metric and let $\delta = 1$.

\begin{proposition}\label{prop:cognitive-costs}
    For any $\Pi\in\Pi^*$ and $\Pi'\in\Pi^{*+}$,
    \[
        W(\Pi')\geq W(\Pi)\geq W(\Pi_\Omega).
    \]
    With the right inequality strict if $\Pi \ne \Pi_\Omega$.
\end{proposition}
\begin{proof}
    $D(P_\Pi) = D(P_{\Pi'}) =a^*$ by the definitions of $\Pi^*$ and $\Pi^{*+}$. Since $\mathcal R^S_\delta\subseteq\mathcal R^S_{\mathrm{final}}$, maximizing $S$ gives $c(\Pi')\leq c(\Pi)\leq c(\Pi_\Omega)$. Subtracting these costs from  $E_{\omega\sim p_{a^*}}[v(\omega)]$ gives the stated inequalities, and strict coarsening strictly reduces cost.
\end{proof}

In this model, partitions in $\Pi^{*+}$ are weakly preferable to $\Pi^*$, since the principal does not benefit intrinsically from the additional information or have other uses for it. It may be preferable to simplify beyond $\Pi^{*+}$ if it saves cognitive costs without making the chosen action too suboptimal. 

If the principal can report differences in expected value, or if this can be backed out from stochastic choices, it is possible to only incentivize refinements that change decisions when their benefit exceeds the additional cognitive cost. If the second opinion mechanism is modified so that its bonus and penalty apply only when $E_{\omega \sim P_{a_{\Pi_2},\Pi_2}}[v(\omega)] - E_{\omega \sim P_{a_{\Pi_1},\Pi_2}}[v(\omega)]>c(\Pi_2)-c(\Pi_1)$, the same logic as the base mechanism applies. 

There may still be partitions that are not refinements of the equilibrium that lead to higher total utility. That illustrates the conceptual limitations of this approach, if we can extract the principal's utility then we can just have agents optimize for that. The main mechanisms of this paper are targeted at situations where the principal's utility cannot be fully specified for a reward signal and/or the principal places value on being informed rather than only choosing the best action at the lowest cognitive cost.

When an accurate model of the principal's cognitive costs is not available, proxies such as the number of outcomes can work nearly as well. In rich real-world settings, almost all possible distinctions between outcomes are irrelevant to almost all decisions. Every simplicity measure supports coarsening over them, so different measures will closely track each other. For example, reducing the number of distinctions between outcomes reduces the number of tokens used to describe them, which reduces compute costs for LLMs.

\subsection{Agent Cognitive Costs}
Agents may also have cognitive costs associated with providing a partition. In the case of AI agents, we can simply scale up rewards until the cognitive costs are inconsequential.

The principal may have to pay costs proportional to the agent's cognitive effort, for example the cost of running an LLM. If this is greater than the principal's costs, then the principal should evaluate $\Pi_\Omega$ directly. Otherwise, the agent simplifying to $\Pi \in \Pi^{*+}$ saves the principal the most value.

Cognitive costs will be higher in training than after convergence to  equilibrium. This is due to multiple rounds of agents providing partitions, and the principal needing to evaluate multiple partitions. After training converges to an equilibrium, only the first agent needs to be queried given an appropriate choice of input.

Using an AI prediction of the principal's decision can reduce costs associated with repeated evaluation. Such a model can be trained on the principal's final decisions and selected partitions where it has high uncertainty.

In simple settings where the principal is unlikely to miss important information, the costs of training may not be worthwhile. Training is primarily beneficial in complex settings where the principal makes consequential decisions. When training is not worthwhile, our prompt-based experiment shows that asking for decision-relevant predictions is still worthwhile.

\end{document}